\documentclass[a4paper,11pt]{article}
\pdfoutput=1
\usepackage{jheppub}
\usepackage[T1]{fontenc}
\usepackage[latin9]{inputenc}
\usepackage{amsmath}
\usepackage{amssymb}
\usepackage{cancel}
\usepackage{subcaption}
\usepackage{comment}
\usepackage{slashed}
\usepackage{physics}
\usepackage{tikz-cd}

\numberwithin{equation}{section}

\newcommand\beq{\begin{equation}}
\newcommand\eeq{\end{equation}}
\newcommand\N{{\mathcal{N}}}
\newcommand\ly{{\lambda_L}}
\newcommand\mO{{\mathcal{O}}}

\begin{document}
\title{The onset of quantum chaos in disordered CFTs}
\affiliation{Department of Particle Physics and Astrophysics, Weizmann Institute of Science, Rehovot, Israel}
\author[1]{Micha Berkooz,\note{e-mail: {\tt micha.berkooz@weizmann.ac.il}}}
\author[2]{Adar Sharon,\note{e-mail: {\tt adar.sharon@weizmann.ac.il}}}
\author[3]{Navot Silberstein\note{e-mail: {\tt navot.silberstein@weizmann.ac.il}}}
\author[4]{ and Erez Y. Urbach\note{e-mail: {\tt erez.urbach@weizmann.ac.il}}}
\abstract{We study the Lyapunov exponent $\lambda_L$ in quantum field theories with spacetime-independent disorder interactions. Generically $\lambda_L$ can only be computed at isolated points in parameter space, and little is known about the way in which chaos grows as we deform the theory away from weak coupling. In this paper we describe families of theories in which the disorder coupling is an exactly marginal deformation, allowing us to follow $\lambda_L$ from weak to strong coupling. We find surprising behaviors in some cases, including a discontinuous transition into chaos. We also derive self-consistency equations for the two- and four-point functions for products of $N$ nontrivial CFTs deformed by disorder at leading order in $1/N$.}

\maketitle

\section{Introduction}

Disordered theories display an interesting range of physical phenomena. Disorder appears in many experimental setups, but it has also found theoretical applications. 
An especially interesting disordered theory is the SYK model \cite{PhysRevLett.70.3339,KitaevTalk}. The SYK model is defined as taking $N$ free fermions and deforming them by a random interaction:
\begin{equation}\label{eq:SYK_H}
    H=\sum_{i_1,...i_q=1}^N J_{i_1...i_q}\psi_{i_1}...\psi_{i_q}\;,
\end{equation}
with $J_{i_1...i_q}$ a random variable drawn from a Gaussian distribution with zero mean and variance $\langle J_{i_1...i_q}^2\rangle=(q-1)!\frac{J^2}{N^{q-1}}$ (with no sum over repeated indices). 

In addition to the SYK model, there also exist more general SYK-like models which display interesting behaviors and have been studied in detail. These are obtained by taking $N$ free fields $\Psi$ (which can be fermions, bosons or superfields in any dimension) and coupling them using the spacetime-independent disorder interaction\footnote{Throughout this paper, we use only spacetime-independent disorder couplings.}
\begin{equation}\label{eq:SYK_like_interaction_intro}
    \sum_{i_1...i_q}J_{i_1...i_q}\Psi_{i_1}...\Psi_{i_q}\;,
\end{equation}
where $J$ are random variables taken from a Gaussian distribution with zero mean, and the disorder interaction can be either a potential or a superpotential term. For example, taking $\Psi$ to be a free fermion in $0+1$d and interpreting \eqref{eq:SYK_like_interaction_intro} as the Hamiltonian, one obtains the SYK model. Similarly, taking $\Psi$ to be a $1+1$d $\mathcal{N}=(1,1)$ free chiral superfield and interpreting \eqref{eq:SYK_like_interaction_intro} as a superpotential, one obtains the MSW model \cite{Murugan:2017eto}. There are also generalizations to $\mathcal{N}=1$ and $\mathcal{N}=2$ in quantum mechanics \cite{Fu:2016vas}, $1+1$d $\mathcal{N}=(2,0)$ \cite{Peng:2018zap}, $1+1$d $\mathcal{N}=(2,2)$ \cite{Bulycheva:2018qcp}, and $2+1$d $\mathcal{N}=2$ \cite{Chang:2021fmd}, among others.

The disorder allows for some exact computations in these theories in the IR at leading order in $1/N$, assuming that they flow to a scale-invariant fixed point \cite{PhysRevLett.70.3339,Maldacena:2016hyu,Kitaev:2017awl}. In particular, one can write down and solve a Schwinger-Dyson equation for the 2-point function, and solve it using a conformal ansatz. In addition, the diagrams contributing to the four-point function of the theory obey an iterative ladder structure, and so they can be formally resummed, allowing for computation of the full four-point function of the theory. Higher-order correlators can also be computed \cite{Gross:2017aos}.

In this paper we consider a more general construction. Consider a CFT in d dimensions, which contains a primary operator $\mathcal{O}$ of dimension $\Delta$. We will study $N$ copies of this core CFT, deformed by a disorder interaction:
\begin{equation}\label{eq:disordered_CFTs}
    (\text{CFT}^N)+\sum_{i_1...i_q}J_{i_1...i_q}\mathcal{O}_{i_1}...\mathcal{O}_{i_q}\;,
\end{equation}
where $J_{i_1...i_q}$ are again Gaussian random variables with variance $\langle J_{i_1...i_q}^2\rangle= \frac{J^2(q-1)!}{N^{q-1}}$, and the disordered interaction term can be interpreted as a potential or a superpotential. These theories can be studied in conformal perturbation theory in $J$. We will call such theories \textbf{disordered CFTs}. As an example, in this notation we refer to SYK-like models as disordered free fields, since the core CFT is a free field theory.

As discussed above, for disordered free fields there is a simple structure to the diagrammatic contributions to the two- and four-point functions at leading order in $1/N$. Surprisingly, we will be able to show that a similar structure exists also for the computations of two- and four-point functions in general disordered CFTs. For example, one can still write down a Schwinger-Dyson equation for the exact two-point function (see figure \ref{fig:sdeqs}). In addition, the contributions to the four-point function still exhibit an iterative ladder structure (see figure \ref{fig:kernel}), which can be formally resummed. We will write down explicitly the corresponding equations from which the two- and four-point functions can be extracted for general disordered CFTs.

Unfortunately, since interacting CFTs are much more complicated than free theories, the corresponding equations one must solve in order to find the exact two- and four-point functions are much more complicated as well. In particular, in order to solve the equations we must know all $n$-point functions of the CFTs. These computations are thus practically possible only in specific CFTs. In this paper we will be able to use the equations to study disordered generalized free fields and a disordered $\mathcal{N}=(2,2)$ minimal model.  

Although the generalized equations we will write down for the two- and four-point functions for disordered CFTs  are complicated, they still allow us to study a new phenomenon. It is common in the literature to make the disorder deformation relevant, so that the theory flows to some fixed point in the IR. One can also try to use a disorder deformation that is classically marginal for disordered free fields, but these deformations are usually marginally irrelevant (see e.g.~\cite{Berkooz:2017efq}). On the other hand, around nontrivial CFTs there is the possibility that these deformations are exactly marginal. In this paper we will mainly focus on exactly marginal deformations, at least at leading order in $1/N$. In particular, this means that it is not necessary to take $J\to\infty$ in order to find a fixed point; instead, the values of $J$ should parametrize a line of (disordered) fixed points.

The existence of a line of fixed points allows one to ask questions about the $J$-dependence of observables in the theory, rather than just the large-$J$ (or IR) behavior as in disordered free fields. We will mostly by interested in the $J$-dependence of the chaos exponent $\lambda_L(J)$. $\lambda_L$ can be read off from the behavior of an out-of-time-ordered correlator (OTOC) \cite{KitaevTalk2,KitaevTalk,Larkin1969QuasiclassicalMI}, which will be reviewed in detail in this paper. Famously, this chaos exponent is bounded from above $\lambda_L\leq 2\pi/\beta$ \cite{Maldacena:2015waa}, with $\beta$ the inverse temperature. Since we will be working around scale-invariant theories, $\beta$ will be the only scale in the problem and so we will set $\beta=2\pi$ in the following, so that the bound reads $\lambda_L\leq 1$.

Usually, computing the chaos exponent requires the theory to be scale invariant, since this simplifies the computation immensely. As a result, $\lambda_L$ is known mostly for some isolated CFTs. However, the structure described above allows us to compute the chaos exponent as a function of the continuous disorder parameter $J$. We will thus be able to follow $\lambda_L$ from $J=0$, where the theory consists of a product of $N$ decoupled CFTs, to $J\to\infty$, where the theory usually coincides with some SYK-like fixed point. We expect there to be very low chaos at $J=0$ and large chaos at $J=\infty$, and we would like to study how chaos emerges in the theory. For some previous discussions of quantum chaos at weak coupling which are similar to the discussion here, see \cite{Stanford:2015owe,deMelloKoch:2019ywq,Chowdhury:2017jzb,Steinberg:2019uqb,Maldacena:2016hyu,Bulycheva:2018qcp} and references therein, and specifically \cite{Lian:2019axs,Hu:2021hsj} which discussed chaos on a line of fixed points in a non-Lorentz-invariant theory.

The question of how chaos appears following a small deformation of a non-chaotic theory is an extremely complicated one, even in classical systems. Classically, there are diverse types of behaviors that different systems can display, leading to a large range of interesting physical phenomena to study. Specifically, some interesting behavior occurs when one deforms an integrable system away from integrability. An important result in this context is the KAM theorem, which discusses the breakdown of integrability under a small deformation for non-degenerate integrable systems with a finite number of degrees of freedom. Schematically, the KAM theorem states that such systems which are slightly deformed away from integrability still retain a large part of their integrable structure. As a result, a ``large'' deformation away from integrability is required in order to find chaos. The fact that deformations away from integrability retain some integrable structure seems to be more robust than the range implied by the KAM theorem; another example of this behavior beyond the scope of the KAM theorem is the Fermi-Pasta-Ulam-Tsingou problem. 

\begin{figure}
	\centering
	\begin{subfigure}[t]{0.5\textwidth}
		\centering
		\includegraphics[width=0.5\linewidth]{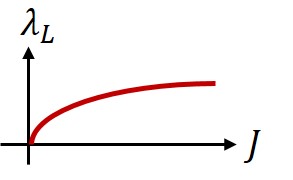}
		\caption{}
		\label{fig:continuous_chaos}
	\end{subfigure}%
	~ 
	\begin{subfigure}[t]{0.5\textwidth}
		\centering
		\includegraphics[width=0.5\linewidth]{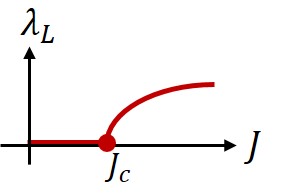}
		\caption{}
		\label{fig:discontinuous_chaos}
	\end{subfigure}
	\caption{The two types of behaviors we find for the dependence of the chaos exponent $\lambda_L$ on the exactly marginal disorder deformation $J$: (a) continuous and (b) discontinuous.}
	\label{fig:dis_continuous_chaos}
\end{figure}

We will find that disordered quantum systems also display different types of behaviors when deformed away from a non-chaotic point. In particular, we will find two types of behavior for the chaos exponent $\lambda_L$, shown in figure \ref{fig:dis_continuous_chaos}.
Schematically, we find that disordered CFTs can either have a continuous or a discontinuous transition into chaos. The discontinuous transition into chaos is an extremely interesting result, and parallels the behavior one might expect from systems which fall under the classes described in KAM theory, but the analogy is very far from being precise. The KAM theorem was an exciting breakthrough in the study of the onset of classical chaos, and we hope that a similar breakthrough may appear in the study of the onset of quantum chaos.

The rest of this paper is organized as follows. In section \ref{sec:chaos_nontrivial_CFT}, we discuss generalities of disordered CFTs at large $N$. We write down the self-consistency equations for the two- and four-point functions, and discuss the computation of the chaos exponent. In section \ref{sec:disorder_and_conformal_manifolds} we define the main classes of theories we will be interested in, where the disorder coupling $J$ is exactly marginal. In section \ref{sec:chaos_GFF} we discuss the first class of theories, the disordered generalized free fields, and we compute the chaos exponent as a function of $J$ for them. In section \ref{sec:min_models_chaos} we discuss the second class of theories, the disordered $\mathcal{N}=2$ minimal models, and we find the chaos exponent in the limit $J\to 0$ for the simplest minimal model. We discuss our results and some future directions in section \ref{sec:conclusions}.

\subsection{Summary of results}

We now summarize the main results of this paper. There are two main results:
\begin{enumerate}
    \item First, we discuss the two- and four-point functions of general disordered CFTs. For the two-point function, we write down a self-consistency equation for the exact propagator (figure \ref{fig:sdeqs}), which generalizes the standard Schwinger-Dyson equations of disordered free fields. We then show that the contributions to the four-point function have an iterative ladder structure (figure \ref{fig:kernel}), again imitating the case for disordered free fields. In both cases the structure is much more complicated in a general CFT compared to a free theory, but it allows for a perturbative expansion in the disorder parameter $J$. 
    \item Second, we discuss chaos in the case where the disorder parameter $J$ is an exactly marginal deformation (at least at leading order in $1/N$). We show that the OTOC also obeys a ladder structure, which allows for a computation of the chaos exponent $\lambda_L(J)$ in the case where $J$ is exactly marginal. We perform this computation in two classes of examples:
    \begin{enumerate}
        \item Disordered generalized free fields: we discuss generalized free fermions in $0+1$d (following \cite{Gross:2017vhb}) and SUSY generalized free chiral superfields in $1+1$d. In both cases $J$ is exactly marginal at leading order in $1/N$. We find that in this class of models there is a discontinuous transition into chaos as in figure \ref{fig:discontinuous_chaos}, so that the chaos exponent vanishes for couplings $J<J_c$ for some finite critical coupling $J_c$, before rising as we raise $J$ above $J_c$. See figure \ref{fig:GRchaos} for the explicit result for the $0+1$d case and figure \ref{fig:2dgffchaos} for the $1+1$d case. Specifically, the chaos exponent read off from the ladder structure appears to be negative for $J<J_c$, but as we explain this just signals a breakdown of some assumptions which are usually made in the computation of the chaos exponent, and should be interpreted as having $\lambda_L=0$ for $J<J_c$.
        \item Disordered SUSY minimal models: we discuss $N$ copies of the $A_{q-1}$ $\mathcal{N}=(2,2)$ minimal models coupled by disorder. The computation of $\lambda_L(J)$ is difficult in general, and so we focus on the particularly simple case of $q=3$, where the theory has central charge $c=1$ and so it reduces to that of a free compact boson. This allows us to compute all possible correlators of the relevant chiral superfields, and to compute the leading contribution to the chaos exponent at small $J$. We find that there is a smooth transition into chaos in this case, as in figure \ref{fig:continuous_chaos}. 
    \end{enumerate}
\end{enumerate}

This paper is just a first step in the study of the onset of quantum chaos in disordered systems, and many open questions remain. In particular, we conjecture a continuity relation around equation \eqref{eq:continuity}, which relates the chaos exponent of disordered CFTs obtained via the retarded kernel in the limit $J\to 0$ to an exponent in a specific limit of a single core CFT. If this is a general result, then it would suffice to study a simple limit of a single copy of the core CFT in order to  find whether the transition into chaos is continuous or discontinuous for the disordered CFTs. It would be very interesting to either prove this result or find a counter-example. Some additional particularly interesting open questions are discussed in the conclusions in section \ref{sec:conclusions}.

Our results are also summarized in the companion paper \cite{Berkooz:2022dfr}.

\section{Disorder around a nontrivial CFT}\label{sec:chaos_nontrivial_CFT}

In disordered free field theories defined in equation \eqref{eq:SYK_like_interaction_intro}, the random interactions impose a specific structure on the perturbative expansion of some observables. This allows for a resummation of Feynman diagrams, and in some cases allows for an exact computation of some observables in the theory at large $N$. In particular:
\begin{itemize}
    \item One can write down a Schwinger-Dyson equation for the two-point function of two $\Psi$'s (which are the free fields at $J=0$). In the conformal limit $J\to \infty$, this can be solved exactly.
    \item In addition, the contributions to the four-point function of $\Psi$'s obey an iterative ladder structure, which can be resummed in principle.
    \item Finally, the out-of-time-ordered four-point function (the OTOC) also has an iterative ladder structure, which can be used to extract the chaos exponent of the theory even without an explicit resummation.
    \end{itemize}

In this section we extend this analysis to disordered CFTs, i.e.~we consider nontrivial (non-free) CFTs which are deformed by a random interaction as in equation \eqref{eq:disordered_CFTs}. Specifically, assume we have a core CFT with some operator $\mathcal{O}$ of dimension $\Delta_{\mathcal{O}}$. Consider a product of $N$ such CFTs, and deform this theory by the interaction
\begin{equation}\label{eq:SYK_like_interaction}
\sum_{i_1\ne i_2 ...\ne i_q}^NJ_{i_1...i_q}\mathcal{O}_{i_1}...\mathcal{O}_{i_q}\;,
\end{equation}
where the indices $\{i_1,...i_N\}$ denote the different CFTs, and $J_{i_1...i_q}$ is again a Gaussian random variable with zero mean and variance $\langle J_{i_1...i_q}^2\rangle= \frac{J^2(q-1)!}{N^{q-1}}$. We take the indices to be different from each other in order to avoid short-distance singularities. The interaction \eqref{eq:SYK_like_interaction} can be a potential term or a superpotential term (in the latter case, all Feynman diagrams that appear in the following should be understood as supergraphs).

Since the core CFT is no longer a free theory, we cannot use the self-consistency equations for the two-and four-point functions discussed above. However, we will show that in the more general case \eqref{eq:SYK_like_interaction}, it is still possible to perform some exact computations in these theories. Specifically, we will show that it is still possible to write down a Schwinger-Dyson (SD) equation for the two-point function, and that the four-point function and the OTOC still obey an iterative ladder structure, which allows one to extract the chaos exponent. We will eventually be interested only in deformations which are exactly marginal, but the self-consistency equations we write will be general. We comment that in general, further counter-terms are required to renormalize the theory, but we will assume that these either average to zero or are not required for the specific correlators we will study (indeed we will see that this is the case for the models discussed in this paper, at least to the order in $J$ we will work in). 

For concreteness, in this section we will assume that the operators $\mathcal{O}$ are real, but the generalization to complex operators is immediate. Similarly, a generalization is immediately possible for chiral superfields $\mathcal{O}_i$ where we interpret \eqref{eq:SYK_like_interaction} as a superpotential, where the diagrams that appear in the following should be understood as supergraphs. 

We begin this section by reviewing how the two- and four-point functions are computed in the standard case of the SYK model. We will then extend this analysis to general disordered CFTs, i.e.~a product of $N$ nontrivial CFTs deformed by the disorder interaction \eqref{eq:SYK_like_interaction}.

\subsection{Review of disordered free fields (the SYK model)}\label{sec:review_of_disordered_free_fields}

We quickly review the computation of the chaos exponent in the SYK model \eqref{eq:SYK_H} \cite{PhysRevLett.70.3339,Maldacena:2016hyu,Kitaev:2017awl}. In the notation described above, the SYK model consists of a core CFT which is a $0+1$d free fermion. We start by writing the Schwinger-Dyson (SD) equations for the two-point functions, followed by the iterative kernel structure for the four-point function. We then discuss the OTOC and the retarded kernel, from which we extract the chaos exponent. This review will later allow us to highlight the similarities and differences between the diagrammatic expansions for disordered free fields and for general disordered CFTs which we will derive later on.

\begin{figure}[]
	\centering
	\includegraphics[width=0.7\linewidth]{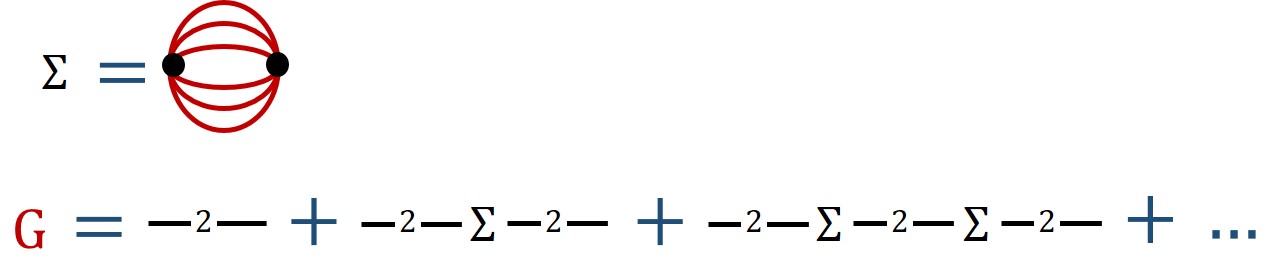}
	\caption{The SD equations for disordered free fields, e.g.~in the SYK model. Red lines correspond to the full two-point function $G$, while ``$-2-$'' corresponds to the two-point function of the core CFT (here a free theory).}
	\label{fig:freesdeqs}
\end{figure}

We start by writing down the SD equations for the two-point function. Consider the computation of the two-point function of two fermions $G(t_1,t_2)=\langle \psi_i(t_1) \psi_i(t_2) \rangle$, where the expectation value $\langle\cdot\rangle$ also averages over couplings. It turns out that the diagrams contributing to the two-point functions at leading order in $1/N$ obey a simple SD equation:
\begin{equation}\label{eq:SYK_SD}
\begin{split}
    G(p)&=\frac{1}{G_0^{-1}(p)-\Sigma(p)}\;,\\
    \Sigma(\tau)&=J^2G(\tau)^{q-1}\;,
    \end{split}
\end{equation}
where $G_0$ is the free fermion propagator, and the second equation defines the self-energy $\Sigma$.
This SD equation is shown diagrammatically in figure \ref{fig:freesdeqs}.

Assuming a conformal theory at $J\to\infty$, one can guess a conformal ansatz for the two-point function of the form 
\begin{equation}
    G(\tau)=b\frac{\text{sgn}(\tau)}{|\tau|^{2\Delta}}\;.
\end{equation}
Plugging this into the SD equations \eqref{eq:SYK_SD}, one finds the solution \cite{Maldacena:2016hyu}
\begin{equation}
    \Delta=1/q,\;\;\;\;\; b^qJ^2=\frac{(1-2\Delta)\tan\pi\Delta}{2\pi}\;.
\end{equation}We thus have the exact two-point function at leading order in $1/N$.

\begin{figure}[]
	\centering
	\includegraphics[width=0.7\linewidth]{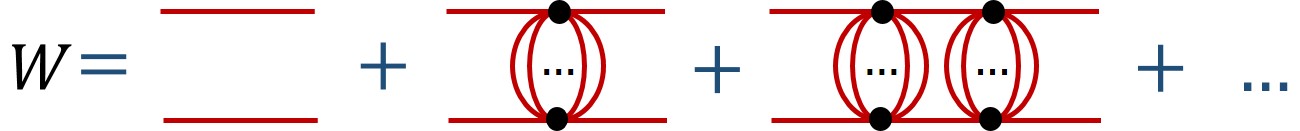}
	\caption{The iterative ladder structure for the four-point function of disordered free fields. Red lines denote full propagators $G$, and black dots denote insertions of the disordered interaction. There are $q-2$ red lines running between each pair of interaction insertions.}
	\label{fig:freeladder}
\end{figure}

We now move on to the four-point function
\begin{equation}
    \sum_{i,j=1}^N\langle \psi_i(t_1)\psi_i(t_2)\psi_j(t_3)\psi_j(t_4) \rangle\;.
\end{equation}
We will be interested in computing the connected contribution of this four-point function, i.e.~we will remove the contribution that is disconnected in the 12 channel.
\begin{equation}
    W=\langle \psi_i(t_1)\psi_i(t_2)\psi_j(t_3)\psi_j(t_4) \rangle_{con} .
\end{equation}
It turns out that contributions to this four-point function follow an iterative ladder structure, see figure \ref{fig:freeladder}.

\begin{figure}[]
	\centering
	\includegraphics[width=0.5\linewidth]{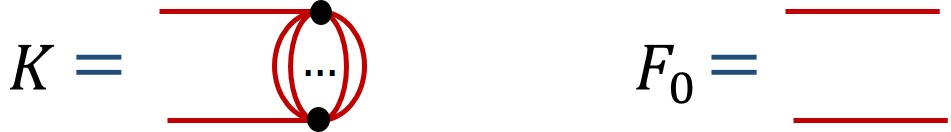}
	\caption{The kernel $K$ and initial contribution $F_0$ for the four-point function for disordered free fields. Red lines again denote full propagators $G$.}
	\label{fig:freeKer}
\end{figure}

Schematically, the ladder structure allows one to write the full four-point function as a geometric series and formally re-sum it
\begin{equation}
    W=\sum_{n=0}^\infty K^n F_0=\frac{F_0}{1-K}\;,
\end{equation}
where $K$ and $F_0$ are defined diagramatically in figure \ref{fig:freeKer}. 
In particular, the spectrum of the theory can be read off from the kernel $K$ by solving the eigenvalue equation $K=1$. The eigenfunctions are constrained by conformal symmetry to be two-point functions of operators of dimension $h$. The eignvalues of $K$ for the SYK model can be computed exactly and were found to be
\begin{equation}
    k(h)=-(q-1)\frac{\psi(\Delta)}{\psi(1-\Delta)}
    \frac{\psi(1-\Delta-h/2)}{\psi(\Delta-h/2)}\;,
\end{equation}
where
\begin{equation}
    \psi(\Delta)=2i\cos(\pi\Delta)\Gamma(1-2\Delta)\;.
\end{equation}

To compute the chaos exponent for the theory, we study the double commutator, defined as\footnote{We will use the terms double-commutator and OTOC interchangeably.}
\begin{equation}\label{eq:OTOC}
\begin{split}
W_R(t_1,t_2) &= \left\langle
[\psi_i(\beta/2),\psi_j(\beta/2+i t_2)] [\psi_i(0),\psi_j(i t_1)]
 \right\rangle_J\\
&=\lim_{\varepsilon\rightarrow0} \left\langle \left( 
\mathcal{\psi}_i\left(\varepsilon\right)-\mathcal{\psi}_i\left(-\varepsilon\right) \right) 
\left(\mathcal{\psi}_i\left(\beta/2+\varepsilon\right)-\mathcal{\psi}_i\left(\beta/2-\varepsilon\right) \right) 
\mathcal{\psi}_j\left(i t_1\right)
\mathcal{\psi}_j\left(\beta/2+i t_2\right) 
\right\rangle_J\;.
\end{split}
\end{equation}
The argument of the operators in the correlator is the Euclidean time. In the second line the operator ordering is assumed to be in increasing Euclidean time. In a chaotic theory, the double-commutator is expected to grow exponentially at large times:
\begin{equation}\label{eq:exponential_growth}
    W_R(t_1,t_2)\sim \exp(\frac{\lambda_L}{2}(t_1+t_2))f(t_1-t_2)\;,
\end{equation}
for some function $f$. $\lambda_L$ is the chaos exponent of the theory. 

Schematically, the appearance of the chaos exponent is due to the out-of-time-ordered (OTO) insertions of operators in the double-commutator.
\begin{figure}[]
	\centering
	\includegraphics[width=0.5\linewidth]{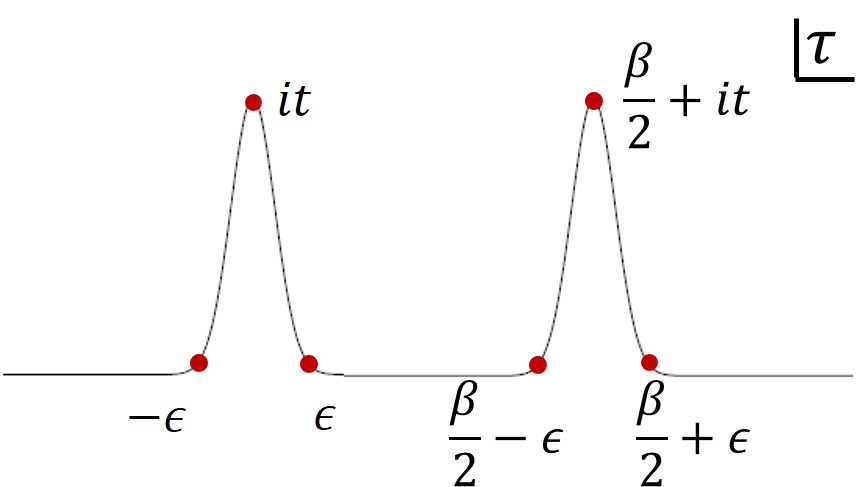}
	\caption{The time contour for the OTOC. Red dots denote all possible positions of operators in the contributions from the double-commutator.}
	\label{fig:time_contour}
\end{figure}
Therefore, by studying the large-time behavior of $W(t_1,t_2)$ we can find the chaos exponent $\lambda_L$. 
In terms of path-integrals, we can account for the different Lorentzian orderings by choosing the complex time contour as in figure \ref{fig:time_contour}. Each insertion of an operator is done at one of the red dots in the figure. As we keep the Euclidean time-ordering (as necessary for convergence), the result is a combination of analytically-continued Euclidean four-point functions $W$. 
As a result, at leading order in $1/N$ $W_R$ has a similar iterative ladder structure to $W$, and we can write down a similar integral equation for $W_R$:
\begin{equation}\label{eq:4pt_Self_consistency}
    W_R=F_{0,R}+K_R W_R\;,
\end{equation}
where $F_{0,R}$ and $K_R$ are specific analytic continuations of $F_0,K$. 
At large times, we assume that the $F_{0,R}$ term is negligible, and so $W_R$ obeys the equation
\begin{equation}
    W_R=K_R W_R
\end{equation}
which is just an eigenvalue equation for $K_R$. We thus find that the exponentially-growing solution for $W_R$ must be an eigenfunction of the retarded kernel $K_R$ with eigenvalue $1$. This allows us to find the chaos exponent $\lambda_L$ by guessing solutions of the form \eqref{eq:exponential_growth} and finding their eigenvalue $k_R(\lambda_L)$ under $K_R$. The largest $\lambda_L$ for which $k_R(\lambda_L)=1$ is the chaos exponent. 

As was the case for the four-point function, the eigenfunctions are constrained by conformal invariance. An eigenfunction of the form
\begin{equation}
    W(t_1,t_2)=\frac{\exp(\lambda(t_1+t_2)/2)}{(2\cosh \frac{1}{2}(t_1-t_2))^{2\Delta+\lambda}}
\end{equation}
has eigenvalue
\begin{equation}\label{eq:SYK_kR_eigenvalues}
    k_{R}(\lambda)=\frac{\Gamma(3-2 \Delta) \Gamma(2 \Delta+\lambda)}{\Gamma(1+2 \Delta) \Gamma(2-2 \Delta+\lambda)}\;.
\end{equation}
In particular, the largest $\lambda$ for which $k_R(\lambda)=1$ is $\lambda_L=1$, i.e.~SYK has maximal chaos.

\subsection{2-point function}\label{sec:2_pt_SD}
\begin{figure}[]
	\centering
	\includegraphics[width=0.75\linewidth]{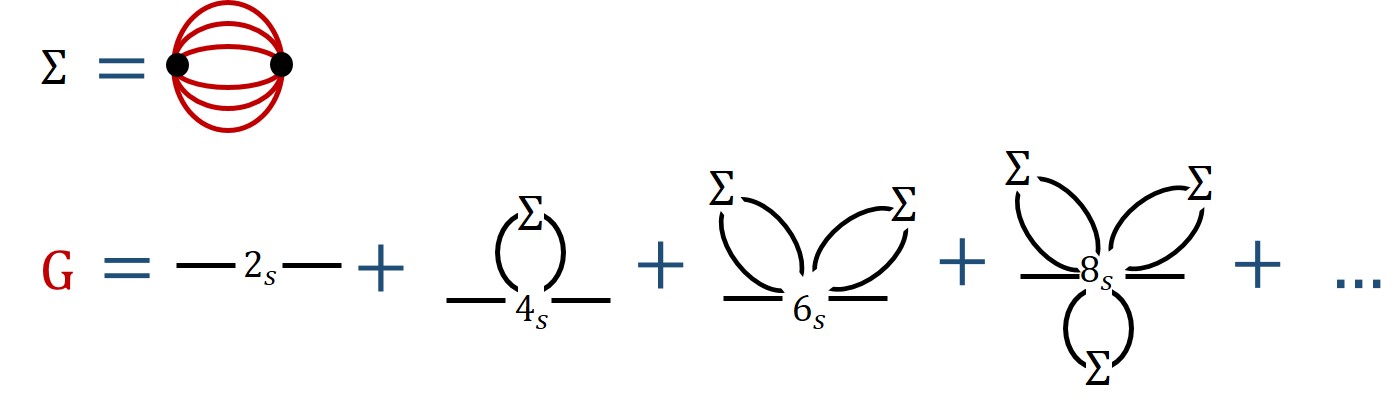}
	\caption{The SD equations. We have emphasized the $G$ insertions using red lines to distinguish them from the 2-point function of the undeformed CFT. Black dots denote insertions of the deformation \eqref{eq:SYK_like_interaction}, and $n_s$ denotes the subtracted $n$-point function defined below.}
	\label{fig:sdeqs}
\end{figure}

\begin{figure}[t]
	\centering
	\includegraphics[width=0.7\linewidth]{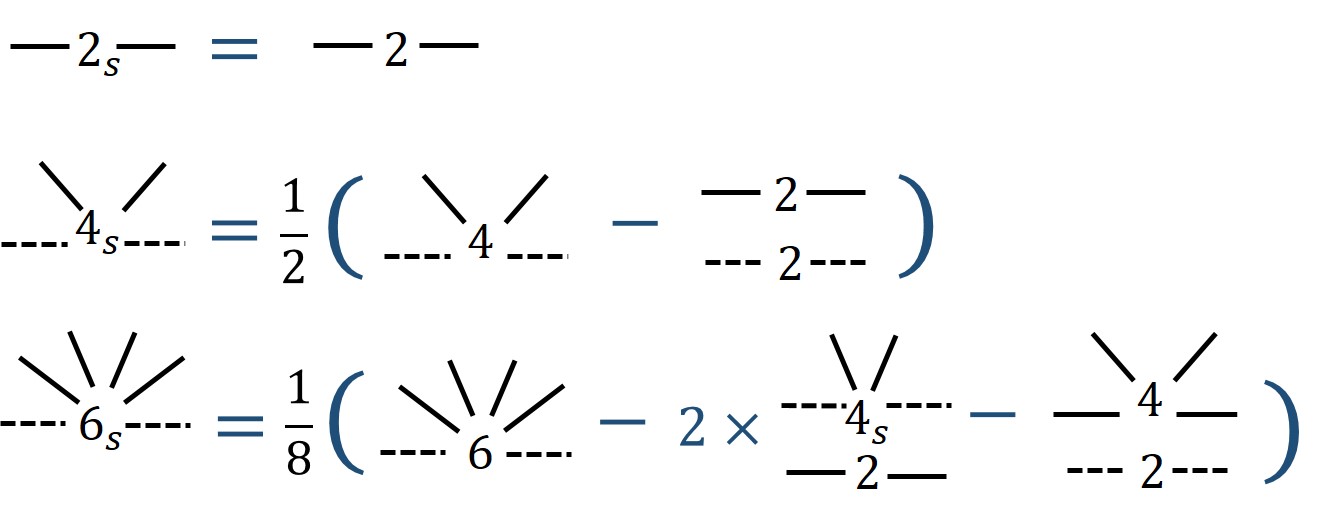}
	\caption{The subtracted $n$-point functions. Dashed lines are connected to  the would-be external positions (see figure \ref{fig:sdeqs}), and the numerical factors indicate symmetry factors. The overall factor for the $n_s$ correlator is in general  $\frac{1}{2^{m}m!}$ where $m=\frac{n}{2}-1$, while the blue numbers inside the bracket correspond to the number of different permutations allowed
	when connecting the legs in a given diagram. The symmetry factors here are for real
	$\mathcal{O}$, but similar expressions exist also for complex fields.}
	\label{fig:subtracted_n_points}
\end{figure}

We now extend the analysis of the SYK model by writing a SD equation for the two-point function for a general disordered CFT, as defined around equation \eqref{eq:SYK_like_interaction}. Starting with a product of $N$ identical core CFTs and adding the deformation \eqref{eq:SYK_like_interaction}, we would like to compute $G(x)=\langle \mathcal{O}_i(x)\mathcal{O}_i(0) \rangle$ at leading order in $1/N$.\footnote{In this section we will assume that $\mO$ is a real field; for complex fields the generalization is straightforward.} This inevitably includes all $n$-point functions of $\mathcal{O}$ at a single core CFT, which we denote without indices: $\langle\mathcal{O}(x_1)...\mathcal{O}(x_n)\rangle$. However, it turns out that there is still an organizing principle for these contributions. We find that $G$ obeys a generalized SD equation which appears in figure \ref{fig:sdeqs} (see also figure \ref{fig:subtracted_n_points}).

In order to derive the SD equations, we follows the standard $G-\Sigma$ formulation of SYK, see for example  \cite{Sarosi:2017ykf}, and we will focus on $0+1$d for concreteness (the generalization to higher dimensions in immediate). Instead of a free theory, we assume some action $S_0[\chi_i]$ for the core theory, for some degrees of freedom $\chi_i$.
The partition function for this theory is\footnote{At leading order in $1/N$ it does not matter whether the disorder is annealed or quenched, so we have assumed it is annealed. In the case where it is quenched the same results can be obtained using the replica trick.}
\begin{equation}
\begin{split}
\langle Z\rangle_J = \int D \chi_i & DG D\Sigma \exp\left( -\sum_iS_0[\chi_i] \right. \\ 
&\left. -\frac{i}{2}\int d\tau d\tau' N\Sigma(G-\sum\frac{1}{N}\mO_i\mO_i) + \frac{J^2N}{2q} \int d\tau d\tau' G(\tau, \tau')^q \right)\;.
\end{split}
\end{equation}
Each operator $\mO_i$ is a local operator of the $i$-th theory, and therefore some function of the $\chi_i$s inside the path-integral.

In the free theory, $S_0$ is quadratic in $\mO$ and so the integral can be calculated. In our case this is not possible. 
Instead, we define  generating functional
\begin{equation} \label{eq:A_def}
A[\Sigma] \equiv \int D\chi \exp( -S_0[\chi] + \frac{1}{2} \int d\tau d\tau' \Sigma \mO(\tau)\mO(\tau') )\;,
\end{equation}
which generates the even $n$-point functions of the undeformed CFT.
The effective action for $G,\Sigma$ is therefore:
\begin{align}
-S[G,\Sigma] =  \frac{N}{2}\left(2\log(A[\Sigma])  + \int d\tau d\tau'\left(\frac{J^2}{q} G(\tau, \tau')^q- \Sigma G\right)\right)\;.
\end{align}
Since $N$ appears as an overall factor in the action, at leading order in $1/N$ the fields $\Sigma, G$ can be evaluated using the saddle point approximation. Varying with respect to $G$ gives the familiar equation $\Sigma=J^2G^{q-1}$. Varying with respect to $\Sigma$ gives:
\begin{equation}\label{eq:G_sol}
G(\tau,\tau')=2\frac{\partial \log(A[\Sigma])}{\partial\Sigma}=2\frac{\partial A/\partial\Sigma}{A}\;.
\end{equation}
The combination of these two equations for $G,\Sigma$ should be considered as the generalization of the SD equation \eqref{eq:SYK_SD} to general disordered CFTs. 

We can rewrite these equations in a form which is more convenient for computations. $A[\Sigma]$ can be expanded as a power series in $\Sigma$. We can thus expand the RHS of \eqref{eq:G_sol} in powers of $\Sigma$, where each order includes a dependence on correlation functions of the core CFT.
Let us compute the first two orders explicitly. At leading order in $\Sigma$, Expanding \eqref{eq:A_def} and plugging it into \eqref{eq:G_sol} gives the expected undeformed two-point function at order $J^0$, $G(\tau,\tau')=\langle\mO(\tau)\mO(\tau')\rangle$. At order $J^2$ (or equivalently, order $\Sigma$) we find
\begin{equation}\label{eq:G_order_J2}
\begin{split}
G(\tau,\tau')|_{J^2}=\int& d\tau_1 d\tau_{2}\Sigma(\tau_1, \tau_{2}) \\
& \cdot \frac{1}{2}\left(\langle\mO(\tau)\mO(\tau')\mO(\tau_1 )\mO(\tau_{2})\rangle-\langle\mO(\tau )\mO(\tau')\rangle\langle\mO(\tau_1 )\mO(\tau_{2})\rangle
\right).
\end{split}
\end{equation}
The first contribution comes from the numerator in \eqref{eq:G_sol} and the second from the denominator. Overall we find a contribution from the four-point function of the undeformed theory, plus an additional subtraction. We call the second line of \eqref{eq:G_order_J2} the ``subtracted four-point function'', and denote it by $4_s$. 
In general, the contribution at order $\Sigma^{n-1}$ will include the $(2n)$-point function, with some additional subtractions involving products of $(2m)$-point functions with $m<n$. We call the full combination the ``subtracted $2n$-point function'', and denote it $(2n)_s$. The explicit form for the subtracted $2,4$ and $6$-point functions are drawn diagrammatically in figure \ref{fig:subtracted_n_points}.
In general $n_s$ will include the $n$-point function with subtractions, times an overall factor of $\frac{1}{2^{n/2-1} (n/2-1)!}$. Overall, we thus find that the SD equations can be written as a sum over contributions from the $(2n)_s$-point functions, as shown in figure \ref{fig:sdeqs}.

It is also useful to understand the subtracted correlators in terms of standard (local) perturbation theory. When computing correlation function in QFT, standard subtractions appear due to the appropriate normalization:
\begin{equation}
\langle \mathcal{O} \rangle=\frac{\int D\phi \mathcal{O}  e^{iS}}{\int D\phi  e^{iS}}\;.
\end{equation}
The denominator subtracts disconnected bubble diagrams from the final result. The result is a sum only over ``connected'' diagrams. This is the same mechanism that requires the subtractions of \eqref{eq:G_sol}. As in standard perturbation theory, this can be used to systematically produce the subtracted correlators. The idea is to write the full $n$-point function in terms of only fully connected pieces, and then subtract the contributions which lead to disconnected diagrams when plugged into the diagrams in figure \ref{app:ns_and_nsp}.
This algorithm is described in detail in appendix \ref{app:ns}.

A simple consistency check of these equations is that they reproduce the expected equations for disordered free fields. We explain how this happens in appendix \ref{app:consistency}. We also emphasize that while a general solution of the equations requires knowing all $n$-point functions of the core CFT, perturbation theory in $J$ to order $J^{n}$ requires only knowing the $(2m)$-point functions for $m\leq n+1$, and so calculations are possible in perturbation theory in $J$.

\subsection{4-point function}\label{sec:four_pt}

\begin{figure}[]
	\centering
	\includegraphics[width=0.75\linewidth]{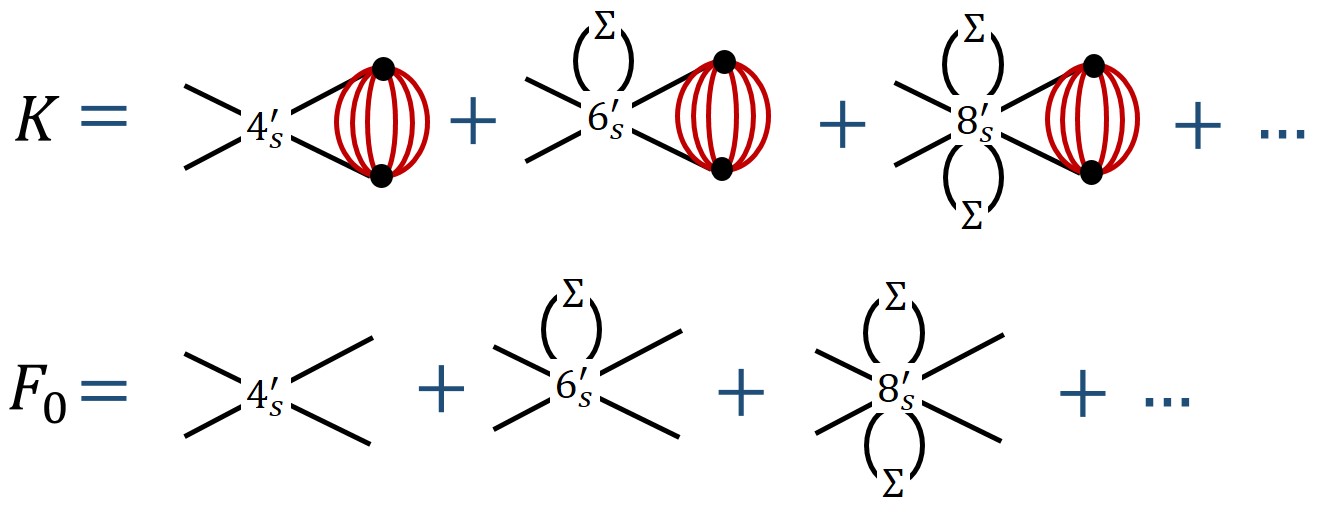}
	\caption{The kernel $K$ and initial diagram $F_0$ for general disordered CFTs. Red lines denote full propagators $G$, and black dots denote insertions of the disorder interaction, with $q-2$ red propagators between each pair.}
	\label{fig:kernel}
\end{figure}

\begin{figure}[]
	\centering
	\includegraphics[width=0.7\linewidth]{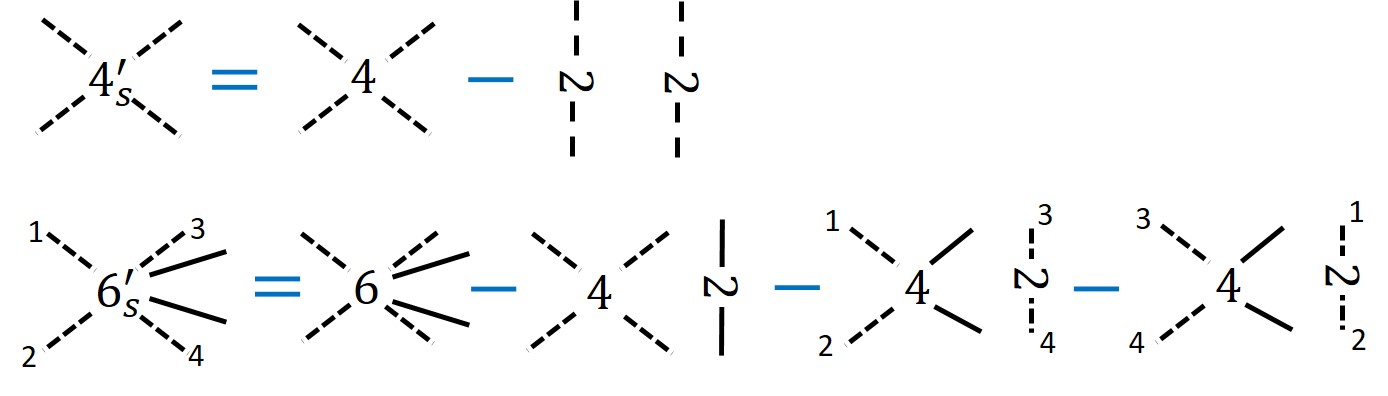}
	\caption{Examples of correlation functions $n_s'$. Dashed lines corresponds to external points, while solid lines are connected via $\Sigma$'s in figure \ref{fig:kernel}.}
	\label{fig:nsprime_examples}
\end{figure}

We now discuss the case of the four-point function. In particular, since we are interested in chaos, we will be considering the connected contribution to the four-point function 
\begin{equation}\label{eq:conn_four}
C=\frac{1}{N^2} \sum_{i,j}\langle\mathcal{O}_i\mathcal{O}_i\mathcal{O}_j\mathcal{O}_j\rangle_{conn}\;,
\end{equation}
where we have suppressed positions. As reviewed in section \ref{sec:review_of_disordered_free_fields}, for disordered free fields one finds a kernel structure. This means that the four-point function obeys
\begin{equation}\label{eq:C_K_F0}
C=\sum_{n=0}^\infty K^n F_0=\frac{F_0}{1-K}\;.
\end{equation}
It turns out that this is also true around a nontrivial CFT. In this case, $K$ and $F_0$ are given diagramatically in figure \ref{fig:kernel}. Here, the correlation functions of the undeformed CFT require slightly different subtractions, and so we have named them $n_s'$. The idea is the same: $n_s'$ is defined by cutting the $n$-point function in every possible way and subtracting the contributions which give disconnected pieces. Explicitly, the first few examples are shown in figure \ref{fig:nsprime_examples}.

We now prove the existence of a kernel structure for the four-point using the $G-\Sigma$ formalism. 
Calculation of the four point function amounts to calculating the two point function of the field $G$.
Following \cite{Sarosi:2017ykf} we expand $G$ and $\Sigma$ to leading order:
$\Sigma = \Sigma_* + |G_*|^{\frac{q-2}{2}}\sigma$ and
$G = G_* + |G_*|^{\frac{2-q}{2}}g$ where $G_*$ is the saddle point solution. The leading correction to the four point is then given by the two point function of the $g$ field. The latter is given at this order by the quadratic piece in the effective action for $g$. The result is that the contribution has an iterative ladder structure
\begin{align}
    C = \langle G(1,2)G(3,4)\rangle - G_*(1,2)G_*(3,4) &= 
    \frac{F_0}{1-K}\;,
\end{align}
where
\begin{equation}\label{eq:four_point_kernel_1}
\begin{split}
F_0(1,2;3,4) &= \frac{\partial^2\log(A[\Sigma])}{\partial\Sigma_{12}\partial\Sigma_{34}}\;,\\
K(1,2;3,4) &=  -4J^2(q-1)\frac{\partial^2\log(A[\Sigma])}{\partial\Sigma_{12}\partial\Sigma_{34}} \cdot G_*^{q-2}(3,4)\;.
\end{split}
\end{equation}

Computations are again more conveniently done after expanding the result in $\Sigma$. First we rewrite the second derivative of $\log[A]$ in a simpler form:
\begin{equation}\label{eq:close_look_at_A}
    4\frac{\partial^2\log(A[\Sigma])}{\partial\Sigma_{12}\partial\Sigma_{34}} = \frac{4}{A}\frac{\partial^2 A[\Sigma]}{\partial\Sigma_{12}\partial\Sigma_{34}} - \frac{4}{A^2}\frac{\partial A[\Sigma]}{\partial\Sigma_{12}}\frac{\partial A[\Sigma]}{\partial\Sigma_{34}}\;.
\end{equation}
The second term in this equation is just the square of the two point function discussed above in \eqref{eq:G_sol}, and is responsible for subtracting diagrams that separate the diagram ``vertically'' (in the sense of figure \ref{fig:kernel}) between points $1,2$ and $3,4$.
Expanding \eqref{eq:close_look_at_A} in $\Sigma$, we again find that we can write $F_0$ and $K$ as a sum over the contributions of new subtracted $n$-point functions which we denote by $n_s'$, see figure \ref{fig:kernel}.

Explicitly, expanding the first term in \eqref{eq:close_look_at_A} in orders in $\Sigma$ one finds
\begin{equation}
\begin{split}
    4\frac{\frac{\delta^2 A[\Sigma]}{\delta\Sigma_{12}\delta\Sigma_{34}}}{A} =& 
    \langle\mathcal{O}(\tau_1)\mathcal{O}(\tau_2)\mathcal{O}(\tau_3)\mathcal{O}(\tau_4)\rangle   \\
    &+\int d\tau_5 d\tau_6 \frac{1}{2}\langle\mathcal{O}(\tau_1)\mathcal{O}(\tau_2)\mathcal{O}(\tau_3)\mathcal{O}(\tau_4)\mathcal{O}(\tau_5)\mathcal{O}(\tau_6)\rangle\Sigma(\tau_5,\tau_6)  \\
    &-\langle \mathcal{O}(\tau_1)\mathcal{O}(\tau_2)\mathcal{O}(\tau_3)\mathcal{O}(\tau_4)\rangle\int d\tau_5 d\tau_6 \frac{1}{2}\langle\mathcal{O}(\tau_5)\mathcal{O}(\tau_6)\rangle\Sigma(\tau_5,\tau_6)\rangle\\
    &+...
    \end{split}
\end{equation}
The first line gives the leading-order contribution, which is just the four-point function. Together with the subtraction from the expansion of $\frac{\partial A[\Sigma]}{\partial\Sigma_{12}}\frac{\partial A[\Sigma]}{\partial\Sigma_{34}}$ term, we find the subtracted $4_s'$ that appears in figure \ref{fig:nsprime_examples}. The next two lines consist of the 6-point contribution, together with a subtraction. Again, combined with the subtraction from the expansion of $\frac{\partial A[\Sigma]}{\partial\Sigma_{12}}\frac{\partial A[\Sigma]}{\partial\Sigma_{34}}$, we find the correlator $6_s'$ in figure \ref{fig:nsprime_examples}. By expanding in orders of $\Sigma$, one can get the exact factors for each diagram at higher orders as well. We give another prescription for finding $n_s'$ in appendix \ref{app:nsprime}. Putting these results together, one finds the series expansion of $F_0$ and $K$ which appears in figure \ref{fig:kernel}.

Once again, it is simple to check that one reproduces the kernel in figure \ref{fig:freeKer} assuming that the undeformed CFT is a free theory. In particular, each $n$-point function amounts to having $n$ $\Sigma$ insertions connecting the $1,2$ points to the $3,4$ points, and resumming these insertions leads to full propagators $G$ connecting the external points, as expected. We also emphasize again that computing $F_0$ and $K$ to a specific order in perturbation theory in $J$ requires knowing only a finite number of correlators in the core CFT.

\subsection{The double-commutator}

In the previous section we found that even for a general CFT deformed by the disorder interaction \ref{eq:SYK_like_interaction}, the four-point function still has a kernel structure. In this section we show that the double-commutator also has a similar structure. As in the SYK model, the double-commutator can be used to measure the chaos exponent, which we will discuss in the next section.

Following the calculation for the SYK model in section \eqref{sec:review_of_disordered_free_fields}, we would like to compute the double commutator:
\begin{equation}\label{eq:OTOC_O}
\begin{split}
W_R&(t_1,t_2) = \frac{1}{N^2}
\sum_{i,j=1}^N \left\langle
[\mO_i(\beta/2),\mO_j(\beta/2+i t_2)] [\mO_i(0),\mO_j(i t_1)]
 \right\rangle_J\\
&=\lim_{\varepsilon\rightarrow0}\frac{1}{N^2}
\sum_{i,j=1}^N \left\langle \left( 
\mO_i\left(\varepsilon\right)-\mO_i\left(-\varepsilon\right) \right) 
\left(\mO_i\left(\beta/2+\varepsilon\right)-\mO_i\left(\beta/2-\varepsilon\right) \right)
\right.\\
& \qquad\qquad\qquad\qquad\qquad\qquad\qquad\qquad\qquad \cdot \left.
\mO_j\left(i t_1\right)
\mO_j\left(\beta/2+i t_2\right) 
\right\rangle_J\;.
\end{split}
\end{equation}
We have again suppressed the spatial coordinates, keeping only the (real and imaginary) time coordinate. By $\langle ... \rangle$ we mean the Euclidean time-ordered thermal trace, where by ``time-ordering'' we mean operator insertions at increasing real (Euclidean) times. The subscript $J$ is meant to emphasize that this is a correlator in the deformed theory \eqref{eq:SYK_like_interaction}.

Note that  \eqref{eq:OTOC_O} is just a combination of analytically-continued  Euclidean (connected) four-point functions on the cylinder. In $0+1$d and $1+1$d, the latter is an analytically-continued flat-space correlator \eqref{eq:conn_four}. For this reason, we will focus for the rest of the paper on $0+1$d and $1+1$d theories. In the previous section we saw that \eqref{eq:conn_four} has a kernel structure. By the same logic, $W_R$ also has the same kernel structure diagramatics, but with the analytically-continued (Euclidean) time contour $\mathcal{C}$ shown in figure \ref{fig:time_contour}. This curve has an increasing Euclidean time (as required by convergence) from $0$ to $\beta$, but is deformed to go through the points $i t$ and $\beta/2 + i t$ over Euclidean time of $2\varepsilon$. These two deformations from the Euclidean integration contour will be called the two ``left/right rails'' respectively. Following the SYK diagramatics, we call each multiplication of the analytically continued version of $K$ from \eqref{eq:C_K_F0} a ``rung''.
Following \cite{KitaevTalk2,Maldacena:2016hyu,Murugan:2017eto}, we can now show by induction how
the complex time integrals simplify into ``retarded'' diagrams in the limit $\varepsilon\rightarrow 0$.

We start with a one-rung first ladder. Denoting $\Delta\mathcal{O}(z)=\mathcal{O}(z+\varepsilon)-\mathcal{O}(z-\varepsilon)$ for brevity, a $(2n)_s'$ ($n\ge 2$) correlator contributes
\begin{equation}
\begin{split}
\int_{\mathcal{C}} dz_5 ...\int_{\mathcal{C}} dz_{2n}
& \left\langle 
\Delta\mathcal{O}(0)
\Delta\mathcal{O}(\beta/2)
\mathcal{O}\left(z_3 \right)\mathcal{O}\left(z_{4}\right)...
\mathcal{O}\left(z_{2n-1}\right)\mathcal{O}\left(z_{2n}\right)\right\rangle^{'}_s  \\
& \cdot \Sigma\left(z_5 ,z_6 \right)\cdot ...\cdot \Sigma\left(z_{2n-1},z_{2n}\right),
\end{split}
\end{equation}
where the $z_i$ are the complex time coordinates along the curve $\mathcal{C}$. $z_3,z_4$ are connected to the next rung and so we do not integrate over them in this (single-rung) expression. Note that the correlator is the (analytically continued) $(2n)_s'$ correlator described in the last section.
The contour integral $\mathcal{C}$ (see figure \ref{fig:time_contour}) is composed of a horizontal Euclidean region between $0$ and $\beta$, and two vertical 'rails' that connects to $z = i t_1$ and $z= \beta/2 + i t_2$~\cite{Murugan:2017eto}.
As a first step we would like to show that the $z_5,...,z_{2n}$ integrals over the rails cancel, and we are left only with the original Euclidean integral for $z_5,...,z_{2n}$. Indeed, assume that several or all of the integration variables are on the rails and assume that $z_k^\pm= i t_k \pm \varepsilon$ is the corresponding coordinate with the largest imaginary time $t_k$. The $z_k$ integral along the rail between $z_k^-$ to $z_k^+$ has the same Euclidean order and so both the $2n$-point function and its corresponding $\Sigma(z_k,z_{k+1})$ will not change, and the integral will cancel between the two directions on the rail. Therefore we can reduce the contour integral for $z_5,...,z_{2n}$ back to the Euclidean time axis $d\tau_5 ... d\tau_{2n}$ between $0$ and $\beta$.

Now we can take the limit $\varepsilon\to 0$ also for the external legs just like in \cite{Murugan:2017eto}.\footnote{In \cite{Murugan:2017eto}, the limit was taken after plugging in explicit expressions for the correlators. Here we will work more generally and show that a similar simplification occurs.} In order for the limits not to vanish we need $z_3$ and $z_4$ to be on different rails.\footnote{In general, we need to allow $z_3,z_4$ to be both in either rail, since in general the $(2n)_s'$-point correlation function  doesn't factorize (as in \cite{Murugan:2017eto}). We will choose $3,4$ to be at the same rail as $1,2$ respectively and account for the other contribution by ``$3\leftrightarrow 4$''.}
We can continue this analysis inductively and get the same result for every rung of the ladder, and in addition we find that the Lorentzian time must increase along the rail. The result is that the double-commutator satisfies the ladder equation
\begin{equation}\label{eq:W_R_eq}
    W_R(t_1,t_2) = F_{0R}(t_1,t_2) + \int_0^{t_1} dt_3 \int_0^{t_2} dt_4 K_R(t_1,t_2,t_3,t_4) W_R(t_3,t_4)\;.
\end{equation}
Here, $F_{0R}$ is the initial contribution, which has a perturbative expansion in $J$ (just like $F_0$ discussed above):
\begin{equation}
\begin{split}
    F_{0R}(t_1,t_2) & = \sum_{n=0}^{\infty} J^{2n} F_{0R}^{(n)}(0,0,t_1,t_2)\\
    F_{0R}^{(n)}(t_1,t_2,t_3,t_4) &= \prod_{i=1}^{2n} \int_0^{\beta} d\tau_i \left\langle \Delta\mathcal{O}\left(it_1 \right)\Delta\mathcal{O}\left(\frac{\beta}{2}+it_{2}\right)\mathcal{O}\left(it_3 \right)\mathcal{O}\left(\frac{\beta}{2}+it_4 \right)
     \prod_{i=1}^{2n}\mathcal{O}\left(\tau_i \right)\right\rangle^\prime_{s}\;,\\
    &\cdot \Sigma\left(\tau_{56}\right)...\Sigma\left(\tau_{2n-1}-\tau_{2n}\right)\;.
\end{split}
\end{equation}
The retarded kernel $K_R$ can also be expanded (again, just like $K$ above):
\begin{equation}
\begin{split}
    K_R(t_1,t_2) & = \sum_{n=0}^{\infty} J^{2n+2} K_R^{(n)}(0,0,t_1,t_2)\\
    K_R^{(n)}(t_1,t_2,t_3,t_4) &= \prod_{i=1}^{2n} \int_0^{\beta} d\tau_i \left\langle \Delta\mathcal{O}\left(it_1 \right)\Delta\mathcal{O}\left(\frac{\beta}{2}+it_{2}\right)\mathcal{O}\left(it_3 \right)\mathcal{O}\left(\frac{\beta}{2}+it_4 \right)
     \prod_{i=1}^{2n}\mathcal{O}\left(\tau_i \right)\right\rangle^\prime_{s}\\
    &\cdot \Sigma\left(\tau_{56}\right)...\Sigma\left(\tau_{2n-1}-\tau_{2n}\right) \cdot G^{q-2}_{lr,\Delta}(3,4)
\end{split}
\end{equation}
In the expression above, the thermal two-point function for a scalar operator of dimension $\Delta$ between points from different rails is
\begin{equation}\label{eq:Glr}
    G_{lr,\Delta}(1,2) = \frac{1}{\left( 4\cosh(\frac{t_{12}-x_{12}}{2})\cosh(\frac{t_{12}+x_{12}}{2}) \right)^\Delta}.
\end{equation}
Because each of the terms in $F_{0R},K_R$ include commutators, the space time events $3,4$ are supported in the (Lorentzian) past of $1,2$ respectively. Note that there is an implicit dependence on the $J$ inside $\Sigma(\tau)$ (see section \ref{sec:2_pt_SD}).

To summarize, we have found that diagramatically, the expression for the double-commutator is similar to that of the four-point function in figure \ref{fig:kernel}, and includes a ladder structure. The difference is that on the two rails, the integrals are Lorentzian, and one should modify the propagators and correlators using a specific analytic continuation. In addition, for the double-commutator the region of integrations is the past of points $1,2$.

\subsection{Chaos}\label{sec:chaos}

\subsubsection{Generalities}

We would like to study the chaos exponent $\lambda_L$, which can be read off of the OTOC $W_R$ at large $t_1,t_2$. As explained in section \ref{sec:review_of_disordered_free_fields}, in the conformal limit of SYK $\lambda_L$ can be found by solving for the eigenvalues $k_R(\lambda)$ of the retarded kernel $K_R$ using an ansatz for the eigenfunctions, and finding values of $\lambda$ for which the eigenvalue is one, $k_R(\lambda)=1$.
For our disordered CFTs, we so far only assumed that the deformation \eqref{eq:SYK_like_interaction} is renormalizable. To study chaos in a similar fashion, we focus in this paper on cases where the deformation \eqref{eq:SYK_like_interaction} is exactly marginal (in the sense that the averaged correlators are conformal for any value of $J$). In fact, in most cases discussed in this paper the coupled theory is conformal for any realization of $J_{i_1 ... i_q}$, and the disordered theory is an ensemble average of CFTs (parameterized by $J$). Since the theory remains conformal at every $J$, and since in addition we saw that the OTOC has a ladder structure even for general disordered CFTs, we can thus follow the same logic as in \ref{sec:review_of_disordered_free_fields} to find the chaos exponent of the theory $\lambda_L(J)$ as a function of $J$ for disordered CFTs. We now explain the method to compute $\lambda_L(J)$ in these theories. For the rest of the section we will focus on $1+1$d theories for concreteness, since most examples discussed in this paper will be in $1+1$d.

The ansatz for the eigenfunctions $W(t_1,t_2)$ of $K_R$ at large times $t_1,t_2 \gg 1$ in the limit where $F_0$ is subdominant is
\begin{equation}\label{eq:W_ansatz}
    W(1,2) = \frac{\exp(-\frac{h+\tilde h}{2}(t_1+t_2)+\frac{h-\tilde h}{2}(x_1+x_2))}
    {(2\cosh(\frac{t_{12}-x_{12}}{2}))^{\Delta-h}
    (2\cosh(\frac{t_{12}+x_{12}}{2}))^{\Delta-\tilde h}}\;.
\end{equation}
As explained in \cite{Murugan:2017eto}, space-normalizability of $W$ require in general $h=-\frac{\lambda}{2} + i p, \tilde h=-\frac{\lambda}{2} - i p$ for real $\lambda, p$. For the kernels presented in \cite{Murugan:2017eto}, as well as the kernels we will find for our explicit examples in sections \ref{sec:chaos_GFF} and \ref{sec:min_models_chaos}, it is possible to show that the minimal solution satisfying $k_R=1$ has $p=0$. We will therefore proceed under the assumption that $p=0$. Setting $h=\tilde h = -\frac{\lambda}{2}$, the eigenvalues are 
\begin{equation} \label{eq:kR_def}
\begin{split}
    k_{R}(\lambda,J) & = \frac{\int d^2 x_3 d^2 x_4 K_R \cdot W}{W} \\
    &= \int d^2 x_3 d^2 x_4 K_R(1,2,3,4) \frac{G_{lr,\Delta+\frac{\lambda}{2}}(3,4)}{G_{lr,\Delta+\frac{\lambda}{2}}(1,2)} e^{\frac{\lambda}{2}(t_3+t_4-t_1-t_2)}\\
    &= \sum_{n=0}^\infty J^{2+2n} 
    \int dx_3 dt_3 dx_4 dt_4 K_R^{(n)}(1,2,3,4)\frac{G_{lr,\Delta+\frac{\lambda}{2}}(3,4)}{G_{lr,\Delta+\frac{\lambda}{2}}(1,2)} e^{\frac{\lambda}{2}(t_3+t_4-t_1-t_2)}\;.
\end{split}
\end{equation}
By conformal invariance, the integral is independent of the $1,2$ coordinates. Note that assuming $t_1,t_2\gg 1$ in the ansatz \eqref{eq:W_ansatz} is equivalent to shifting the integration domain of $t_3,t_4$ to start from $-\infty$ (and not $0$ as in \eqref{eq:W_R_eq}). As a result, the integral over the events $3,4$ is exactly over the past of the events $1,2$ respectively.
As in SYK, the chaos exponent is the largest value of $\lambda_L$ such that $k_R(\lambda_L,J)=1$.\footnote{The $1+1$ dimensional examples we will give below will all be supersymmetric. In this case one can consider different kernels $K_R$ by taking different operators on the various external legs. In practice, in previous studies the dominant contribution to the chaos exponent always comes from the bosonic kernel, i.e. with the bottom components appearing on external legs. We will assume this is the case here as well, and focus on the bosonic kernel. We have checked explicitly for some of the examples to be discussed in the paper that indeed the bosonic kernel gives the most dominant contribution.}

In general, this equation is extremely complicated, and we will only solve it exactly in this paper for the case of disordered generalized free fields. For a more general CFT, one can instead hope to solve it in perturbation theory in $J$, which we discuss next. 

\subsubsection{Chaos at $J\rightarrow 0^+$}\label{sec:chaos_J0}

In this section we will discuss the behavior of $\lambda_L(J)$ close to $J=0$. We start by discussing chaos at $J=0$, and then discuss chaos at small finite $J$. 

First we consider the strict limit of $J=0$, where the $N$ CFTs decouple and the kernel vanishes. At $J=0$, the late-time behavior of the double-commutator $W_R$ defined in \eqref{eq:OTOC_O} is given by
\begin{equation} \label{eq:lambdaL0_def}
\begin{split}
    \lim_{t_1,t_2\rightarrow \infty} W_R(t_1,t_2) \mid_{J=0} & = \lim_{t_1,t_2\rightarrow \infty} \frac{1}{N} \langle [O(\beta/2),O(\beta/2+i t_2)][O(0),O(i t_1)]\rangle \\
    & \equiv \frac{1}{N} \exp(+\lambda_L^0 (t_1+t_2)/2)\;.
\end{split}
\end{equation}
The first equality is a result of the decoupling at $J=0$, and the second equality is a definition: the large $t_1,t_2$ behavior of the undeformed retarded 4-point function is controlled by an exponent which we denote $\lambda_L^0$.\footnote{As shown in \cite{Murugan:2017eto}, in $1+1$d this behavior is equivalent to the Regge behavior of the four-point function. We suspect the discussion of this section is parallel to previous studies of the Regge trajectory in weak coupling \cite{Costa:2012cb}.}

Despite its name, $\lambda_L^0$ is not the chaos exponent of a single undeformed CFT. The reason is that in \eqref{eq:lambdaL0_def} we take $t_1,t_2$ to be larger then the scrambling time of the undeformed theory. Instead, \eqref{eq:lambdaL0_def} simply studies the eventual decay of the double-commutator, a decay that is required by unitarity (see \cite{Caron-Huot:2017vep} for a recent discussion). As a result, even if a single undeformed CFT is chaotic, we expect this exponent to be non-positive: $\lambda_L^0 \le 0$.
Nevertheless, in this section we will try to motivate physically (and later verify in the examples below) why $\lambda_L(J)$ is a continuous function at $J=0$, in the following sense: for every $J>0$ one can calculate $\lambda_L(J)$ by solving the eigenvalue equation $k_R(\lambda,J)=1$ in \eqref{eq:kR_def}, and in the limit $J\to 0^+$ this coincides with the result for the decoupled theory, i.e.~
\begin{equation}\label{eq:continuity}
    \lambda_L(J\to 0^+)=\lambda_L^0\;.
\end{equation}
We stress that this is not a direct outcome of standard conformal perturbation theory, since even the leading order of the kernel appears in infinitely many diagrams in the four-point $W$, and therefore might lead to nontrivial late-time behavior. Our argument doesn't rule out further discontinuities in $\lambda_L(J)$ for $J>0$, although in known examples no such discontinuities were found.

Moving on the nonzero $J$, the chaos exponent $\lambda_L(J)$ is given by solving the eigenvalue equation $k_R(\lambda)=1$ with $k_R$ as in \eqref{eq:kR_def}. We would like to study the exponent at small nonzero $J$ by solving the eigenvalue equation $k_R=1$ in this limit. As the operator $K_R$ can be expressed in orders of $J$, one can compute its eigenvalues $k_R(\lambda)$ in orders of $J$. 
As a first step, we assume the equation $k_R=1$ can be solved in orders of $J$, and focus on the leading term in \eqref{eq:kR_def}:
\begin{equation}\label{eq:kR_4p}
\begin{split}
    k_R(\lambda,J)
    =& J^2 \int d^2 x_3 d^2 x_4 \left\langle \Delta\mathcal{O}\left(it_1 \right)\Delta\mathcal{O}\left(\beta/2+it_{2}\right)\mathcal{O}\left(it_3 \right)\mathcal{O}\left(\beta/2+it_4 \right) \right\rangle'_s \\
    & \cdot \frac{G_{lr,\Delta+\frac{\lambda}{2}}(3,4)}{G_{lr,\Delta+\frac{\lambda}{2}}(1,2)} \exp\left(\frac{\lambda}{2} (t_3+t_4-t_1-t_2)\right)
    \cdot G_{lr,2\Delta(q-2)}(3,4) +O(J^4)\\
      =& \frac{J^2}{4} \frac{\exp(-\frac{\lambda}{2} (t_1+t_2))}{G_{lr,\lambda/2}(1,2)} 
     \frac{\int^\infty_{u_1} du_3 \int_{-\infty}^{u_2} du_4}{u_{34}^{2+\frac{\lambda}{2}}}
     \frac{\int_{v_1}^\infty dv_3 \int_{-\infty}^{v_2} dv_4}{v_{34}^{2+\frac{\lambda}{2}}} \mathcal{G}_R(\chi,\bar\chi)  +O(J^4)\;.
\end{split}
\end{equation}
Here we changed variables to
\begin{equation}
    \begin{split}
        u_3=e^{x_3-t_3}\;,&\quad v_3=e^{-x_3-t_3}\;,\\
        u_4=-e^{x_4-t_4}\;,&\quad v_4=-e^{-x_4-t_4}\;.
    \end{split}
\end{equation}
In these variables the conformal ratios are 
\begin{equation}
    \chi=\frac{u_{12} u_{34}}{u_{14}u_{32}},\quad  \overline{\chi}=\frac{v_{12} v_{34}}{v_{14}v_{32}}\;. 
\end{equation}
$\mathcal{G}_R$ is the retarded normalized 4-point of the undeformed CFT at $J=0$
\begin{equation}\label{eq:def_G_R}
\begin{split}
    \mathcal{G}_R(\chi,\bar\chi) &= \frac{\left\langle  
    [O(\beta/2+it_2),O(\beta/2+it_4)]
    [O(it_1),O(it_3)]
    \right\rangle_s'}{G_{lr,\Delta}(1,2)G_{lr,\Delta}(3,4)}\\
    & = \lim_{\varepsilon_1,\varepsilon_1\rightarrow 0} \mathcal{G}_{++}-\mathcal{G}_{+-}-\mathcal{G}_{-+}+\mathcal{G}_{--}
\end{split}
\end{equation}
with the normalized four-point $\mathcal{G}_{\pm_1,\pm_2} = \mathcal{G}(u_1 e^{\pm i \varepsilon_1}, v_1 e^{\pm i \varepsilon_1},u_2 e^{\pm i \varepsilon_2}, v_1 e^{\pm i \varepsilon_2},u_3,v_3,u_4,v_4)$.
Similar expressions can be written at any order of $J$, where at order $J^{2n}$ each integral will now include $n-1$ Euclidean flat space integrations. Note that at this order in $J$, only the four-point function contributes to the eigenvalue.

Importantly, we find that as $J\to 0^+$, the eigenvalue \eqref{eq:kR_4p} vanishes since it is proportional to $J^2$. As a result, the only way to have $k_R=1$ as $J$ approaches zero (at this order) is for $\lambda$ to also approach a value at which the integral diverges like $1/J^2$. Thus the chaos exponent in the limit $\lambda_L(J\to0^+)$ is found by looking for values of $\lambda$ for which the integral in \eqref{eq:kR_4p} diverges.

How can the integral \eqref{eq:kR_4p} acquire divergences as a function of $\lambda$?
The only short-distance singularities that may appear are when $x_3,x_4$ approach $x_1,x_2$ respectively, but such a divergence will appear independently of $\lambda$.\footnote{Other pairings of the $x_i$-s won't develop short distance singularities due to the euclidean separation (they are on different rails).}
Furthermore, the integral itself should converge for large enough values of $\lambda$. The only possible range for a divergence is thus the combined limit of $|u_3|,|u_4|\rightarrow \infty$ (or $|v_3|,|v_4|\rightarrow \infty$, or both), corresponding to taking the two points to be very far in the past. As explained near \eqref{eq:lambdaL0_def}, in this limit we expect the retarded four-point function to behave exponentially due to the chaotic properties of the original (undeformed) CFT. 
The rest of the integrand in \eqref{eq:kR_4p} is exponentially decaying in this limit as $\exp(\frac{\lambda}{2}(t_3+t_4))$ ($t_3,t_4$ approach $-\infty$). Combining this with \eqref{eq:lambdaL0_def}, the total exponent of the integrand is $\lambda_L^0-\lambda$.
We therefore expect the integral to diverge exactly for $\lambda\le\lambda_L^0$, although a more rigorous argument will require a more careful analysis of the integral.
If indeed the integral diverges only for $\lambda \le \lambda_L^0$, the result is that at least to leading order, the solution to $k_R\mid_{J^2}=1$ in the limit $J\to 0^+$ is $\lambda_L^0$. In other words, $\lambda_L(J\to0^+)=\lambda_L^0$, and the continuity relation \eqref{eq:continuity} is obeyed.

What do we expect to get at higher orders of $J$? Consistency requires that higher-order contributions to $k_R$ also remain finite for $\lambda>\lambda_L^0$. 
Indeed, we can argue on physical grounds that the integrals appearing at higher orders diverge only for $\lambda\le\lambda_L^0$. To this end, we ask how the higher order integrals can diverge.
Higher order contributions to $k_R$ are similar to \eqref{eq:kR_4p}, only with extra Euclidean insertions (together with a finite Euclidean integral). After taking care of all the normal Euclidean short-distance singularities we again expect an exponential behavior as we send the points $3,4$ to the past together. It is reasonable to assume the late-time behavior of a single copy of the core CFT is controlled by a single exponent, the same one that controlled the double-commutator late-time behavior \eqref{eq:lambdaL0_def}. We are therefore assuming, in accordance with \eqref{eq:lambdaL0_def} that the double commutator together with the other euclidean insertions satisfies
\begin{equation}\label{eq:ansatz}
    \lim_{t_1,t_2\rightarrow \infty} \langle [O(\beta/2),O(\beta/2+i t_2)] [O(0),O(i t_1)] \mathcal{O}\left(\tau_1\right)...\mathcal{O}\left(\tau_n\right)]\rangle =e^{+\lambda_L^0 (t_1+t_2)/2}\cdot f_n(\tau_i)\;,
\end{equation}
for some functions $f_n(\tau_i)$. As a result, we expect higher order integrals to diverge only for $\lambda \le \lambda_L^0$, and our perturbative expansion of the retarded kernel was justified. The result is that the full kernel $k_R(\lambda)$ diverges as $\lambda\to \lambda_L^0$ at finite $J$, and as we take $J\to 0^+$ the solution to the full eigenvalue equation would approach $\lambda_L(J\to0^+)\to \lambda_L^0$.
We will explicitly show that the assumption \eqref{eq:ansatz} is justified for all examples discussed in this paper.

\subsubsection{A toy model}

In order to illustrate these points, we now provide a simple example of computing the chaos exponent at $J\to 0^+$ for a simple toy model which has many similarities to the explicit theories we will consider in this paper. 

Consider a normalized retarded four-point function of the form $\mathcal{G}_R(\chi,\bar\chi) = (\chi \bar\chi)^{-\lambda^0_L}$, where $\chi$ is the usual conformal cross-ratio. This function satisfies \eqref{eq:lambdaL0_def}. One can think of it as the four-point function of a generalized free field, or as the leading term of some more general four-point function in the limit $\chi,\bar\chi\rightarrow 0$.\footnote{As we are discussing the retarded four-point, this is actually the $\chi,\bar\chi\rightarrow 0$ limit in the ``second sheet'', see \cite{Murugan:2017eto}.}
Since we expect the divergence in $\lambda$ to come from the region $\chi,\bar\chi\rightarrow 0$, this should be enough to study chaos.

Substituting $\mathcal{G}_R$ in \eqref{eq:kR_4p} together with $u_1=v_1=1,u_2=v_2=0$ gives
\begin{equation}
\begin{split}
    k_R(\lambda,J) &= \frac{J^2}{4} \left( \int_1^\infty dz_3 \int_{-\infty}^0 dz_4 \frac{1}{|z_{34}|^{2+\frac{\lambda}{2}+\lambda_L^0}|z_3|^{-\lambda_L^0}|z_4-1|^{-\lambda_0} }\right)^2\\
    & = \frac{J^2}{4} \left(\frac{
    \Gamma\left(\frac{\lambda-\lambda_L^0}{2}\right)
    \Gamma^2\left(1+\frac{\lambda_L^0}{2}\right)}
    {\Gamma\left(2+\frac{\lambda+\lambda_L^0}{2}\right)}\right)^2\;.
\end{split}
\end{equation}
The integrals $z_3,z_4$ converge as long as $-2<\lambda_L^0<\lambda$. As a function of $\lambda$, the integral diverges for $\lambda<\lambda_L^0$, and as a result $k_R$ has a double pole at $\lambda=\lambda_L^0$. For a more general $\mathcal{G}_R$ we expect a similar behavior only at small enough $\chi,\bar\chi \ll 1$. Nevertheless we don't expect higher orders in $\chi,\bar\chi$ to add further singularities for $\lambda>\lambda_L^0$. In other words, we expect the general behavior
\begin{equation} \label{eq:chaos_guess}
    k_R(\lambda,J) =  \frac{J^2}{(\lambda-\lambda^0_L)^2}(...) +O(J^4)\;,
\end{equation}
where the ellipses denote an analytic function of $\lambda$ with no further singularities for $\lambda>\lambda^0_L$. For this general behavior we can now find the chaos exponent at leading order in $J$ by solving $k_R(\lambda_L,J)=1$ using \eqref{eq:chaos_guess}. The result is indeed $\lambda_L = \lambda_L^0 + O(J)$. 

For consistency, we must now check contributions from higher $n$-point functions as well. We assume that these also contribute at most a double pole at $\lambda=\lambda_L^0$, due to the same logic described above equation \eqref{eq:ansatz}. Under these assumptions, we can resum the series and find for the eigenvalue \eqref{eq:kR_def}
\begin{equation} \label{eq:k_R_full_guess}
    k_R(\lambda,J) = \frac{J^2}{(\lambda-\lambda_L)^2} f(\lambda,J^2)\;,
\end{equation}
For some $f(\lambda,J^2)$ an analytic function in $\lambda,J^2$ (for $\lambda>\lambda_L^0$). In this form it is evident that the solution to $k_R(\lambda_L,J)=1$ as $J\to 0^+$ is $\lambda_L = \lambda_L^0 + O(J)$. Note that the zeroth order in $J$ requires knowledge of all of the correlators of the undeformed CFT, but higher orders in $\lambda_L$ can be found using finitely many correlators. Specifically, the first order correction depends only on the undeformed four-point function \eqref{eq:kR_4p}. 

While this example only represents a toy model, we will see below that the specific theories we consider match exactly this behavior.

\subsubsection{Summary}\label{sec:chaos_summary}

To summarize, whenever $J$ is exactly marginal, we would like to solve the eigenvalue equations $k_R(\lambda,J)=1$ at weak coupling in perturbation theory in $J$. The leading order is controlled by the largest value of $\lambda$ for which the integral \eqref{eq:kR_def} diverges. For consistency, higher-order integrals must not diverge for higher values of $\lambda$, and we motivated why this would be the case for physical models (and we will check that this is the case explicitly for the models in this paper). It is then possible to find the chaos exponent $\lambda_L(J)$ in perturbation theory in $J$. 

In particular, in the limit $J\to 0^+$, we argued that the divergences in the integrals are determined by the exponent $\lambda_L^0$ of the core CFT, see equation \eqref{eq:lambdaL0_def}. As a result, $\lambda_L(J)$ calculated via the kernel and extrapolated to $J=0$ should coincide with the late-time behavior of the undeformed CFT. In other words, the chaos exponent obeys a ``continuity relation'' \eqref{eq:continuity}, $\lambda_L(J=0^+)=\lambda_L^0$. If this is true, it allows us to compute $\lambda_L(J)$ at small $J$ by performing a simple calculation in a single copy of our core CFT. We will check explicitly in the examples in this paper that the continuity relation is satisfied.

An important comment is in order. As mentioned above, $\lambda_L^0$ does not describe any chaotic behavior. Specifically, it is always non-positive ($\lambda_L^0 \le 0$) from unitarity \cite{Caron-Huot:2017vep}. Consider a case where it is strictly negative, $\lambda_L^0<0$. According to the continuity conjecture \eqref{eq:continuity}, at small enough coupling $\lambda_L(J)$ will be negative as well, $\lambda_L(J) <0$. This result must be explained, since a negative chaos exponent seems unphysical. Indeed, we do not expect $\lambda_L<0$ to be consistent, since in our analysis we neglected terms which are not exponentially growing at large times, but may grow larger than the contribution of the negative chaos exponent. However, if one finds $\lambda_L<0$ using the method above, then at the very least it is clear that there are no solutions to the eigenvalue equation with $\lambda_L>0$, since otherwise they would have appeared. As a result, our interpretation of the case $\lambda_L<0$ is that there is no chaos, and instead the physically sensible solution is $\lambda_L=0$. In other words, the physical chaos exponent is given by $\max(0,\lambda_L)$. As a result, if $\lambda_L<0$ for small $J$, then we expect a discontinuous transition into chaos as in figure \ref{fig:discontinuous_chaos}. On the other hand, if $\lambda_L^0=0$, then the continuity relation \eqref{eq:continuity} leads us to expect $\lambda_L(J) \sim J^2 +O(J^4)$ and a continuous transition into chaos, see figure \ref{fig:continuous_chaos}. In particular, assuming the continuity relation is a general result for disordered CFTs, it allows to distinguish between continuous and discontinuous transitions into chaos by examining the late-time behavior of a single core CFT.


\section{Disorder and Conformal Manifolds}\label{sec:disorder_and_conformal_manifolds}

In this section we introduce examples where a disordered interaction is exactly marginal (at least at leading order in $1/N$). We will discuss two main classes of models: generalized free fields and the $1+1$d $\N=(2,2)$ minimal models. We will also discuss some relations between these two models. In the next sections we will study these models more carefully.

An interesting point is that in the SUSY theories discussed below, we expect the theory to be conformal for every realization of the couplings $J_{i_1...i_q}$, and not just after the average over couplings. Thus these theories are similar to recent discussions of averaging over CFTs, see e.g.~\cite{Maloney:2020nni,Afkhami-Jeddi:2020ezh}. Note that this is not enough in order for the averaged correlators for the operators $\mathcal{O}_i$ in \eqref{eq:SYK_like_interaction}  to still be conformal - it is crucial that in addition, the dimension of the operators $\mathcal{O}_i$ is independent of $J$. Indeed this will be the case since $\mathcal{O}_i$ will be chiral superfields and so their dimension will be protected.

\subsection{Disordered generalized free fields}

\begin{figure}[]
	\centering
	\includegraphics[width=0.5\linewidth]{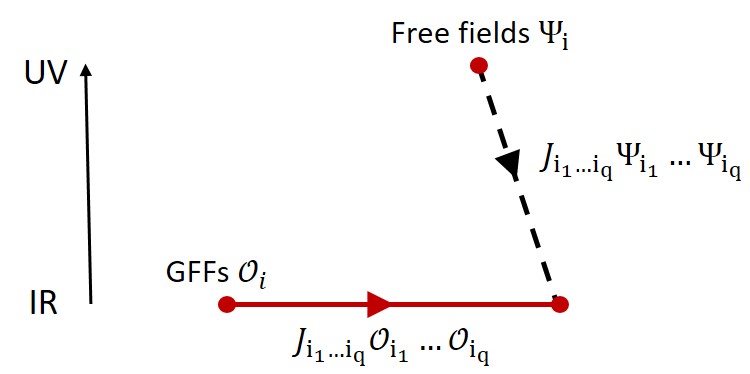}
	\caption{Schematic RG flows for disordered GFFs. Red denotes scale invariant theories. The disorder deformation is exactly marginal, leading to a line of fixed points which ends at the same point obtained by deforming free fields by the same interaction.}
	\label{fig:GFFs_RG}
\end{figure}

We will discuss two main examples of disorder around generalized free fields (GFFs). One is the cSYK line of fixed points in quantum mechanics \cite{Gross:2017vhb}, and another is the $1+1$d $\mathcal{N}=(2,2)$ version of this line of fixed points. The RG flows for these theories are schematically shown in figure \ref{fig:GFFs_RG}. In the figure, we have assumed that disordered GFFs at large $J$ reach the same fixed point as the corresponding disordered free fields at large $J$. Intuitively, this is because the disorder interaction ``dominates'' at large $J$, and so the original core CFT plays no role. Indeed, \cite{Gross:2017vhb} found evidence for this result, and we will provide additional evidence here for both the QM model and the $1+1$d model.

\subsubsection{The cSYK model}\label{sec:cSYK}

Gross and Rosenhaus \cite{Gross:2017vhb} considered a QM theory of $N$ generalized free fermions  coupled via an SYK-like interaction:
\begin{equation}
S=S_0+S_\text{SYK}
\end{equation}
with
\begin{equation}
\begin{split}
S_0&=-\Delta \sum_{i=1}^n  \int d \tau_1 d \tau_2 \chi_i\left(\tau_1\right) \frac{\operatorname{sgn}\left(\tau_1-\tau_2\right)}{\left|\tau_1-\tau_2\right|^{2-2 \Delta}} \chi_i\left(\tau_2\right)\;,\\
S_\text{SYK}&=\frac{i^{\frac{q}{2}}}{q !} \sum_{i_1, \ldots, i_{q}=1}^n  \int d \tau J_{i_1 i_2 \ldots i_{q}} \chi_{i_1} \chi_{i_2} \cdots \chi_{i_{q}}\;.
\end{split}
\end{equation}
Here, $J_{i_1...i_q}$ are Gaussian random variables with variance $\langle J_{i_1...i_q}^2\rangle= \frac{J^2(q-1)!}{N^{q-1}}$. Choosing $\Delta=1/q$, the deformation becomes classically marginal, and it is argued in \cite{Gross:2017vhb} that it is exactly marginal at leading order in $1/N$. Following \cite{Gross:2017vhb} we will call this theory cSYK (``conformal SYK'').

Let us review the computation of the disorder averaged two- and four-point functions in this theory, as a function of the exactly marginal parameter $J$.  For small $J$, we find the propagator of a GFF:
\begin{equation}
G_{UV}=\frac12 \frac{\text{sign}\tau}{|\tau|^{2\Delta}}\;.
\end{equation}
For large $J$, the interaction term dominates and we expect to find the same propagator as in the SYK model. For general $J$, the solution to the SD equations takes the form
\begin{equation}\label{eq:cSYK_prop}
G(\tau)=\frac{\bar b(J) \text{sign}\tau}{|\tau|^{2\Delta}}
\end{equation}
where $\bar b$ solves the equation\footnote{Equation \eqref{eq:bbar} fixes a minus sign that is missing from equation (3.12) of \cite{Gross:2017vhb}.}
\begin{equation}\label{eq:bbar}
\frac{\bar{b}^{q}}{1-2 \bar{b}}=-\frac{1}{J^2 \psi(1-\Delta) \psi(\Delta)}\;\;\;\;\;\;\; \psi(\Delta) \equiv 2 i \cos (\pi \Delta) \Gamma(1-2 \Delta).
\end{equation}
As a consistency check, as $J\to 0$ the solution approaches $\bar b =1/2$ which is the GFF solution, and as $J\to \infty$ the solution approaches the SYK solution (see figure \ref{fig:GFFs_RG}), given by
\begin{equation}
\bar{b}^q(J\to \infty)=-\frac{1}{J^2 \psi(1-\Delta) \psi(\Delta)}\;\;\;\;\;\;\; \psi(\Delta) \equiv 2 i \cos (\pi \Delta) \Gamma(1-2 \Delta)\;.
\end{equation}

Next we discuss the four-point function. The kernel for the SYK model $K_\text{SYK}$ is a product of propagators (see figure \ref{fig:freeKer}):
\begin{equation}
K_\text{SYK}(\tau_1,\tau_2;\tau_3,\tau_4)\propto J^2G(\tau_{13})G(\tau_{24})G(\tau_{34})^{q-2}\;.
\end{equation}
The only effect of considering cSYK is that the propagators change by an overall factor \eqref{eq:cSYK_prop}, and so it is no surprise that the kernel $K_\text{cSYK}$ is the same as $K_\text{SYK}$ up to an overall factor:
\begin{equation}\label{eq:K_cSYK}
\frac{K_\text{cSYK}}{K_\text{SYK}}=\left(\frac{\bar{b}(J)}{\bar{b}(J\to \infty)}\right)^{q}\;.
\end{equation}
Thus the eigenvalues of the cSYK kernel are
\begin{equation}
k_\text{cSYK}(h)=\left(\frac{\bar{b} (J)}{\bar b(J\to \infty)}\right)^{q}k_\text{SYK}=-\left(\frac{\bar{b} (J)}{\bar b(J\to \infty)}\right)^{q} (q-1) \frac{\psi(\Delta)}{\psi(1-\Delta)} \frac{\psi\left(1-\Delta-\frac{h}{2}\right)}{\psi\left(\Delta-\frac{h}{2}\right)}\;.
\end{equation}

\subsubsection{$1+1$d $\mathcal{N}=2$ disordered GFFs}\label{sec:2d_GFFs}

We now describe a $1+1$d SUSY version of the Gross-Rosenhaus model. In other words, we will discuss $1+1$d $\mathcal{N}=2$ GFFs deformed by an SYK-like interaction.\footnote{The $1+1$d $\mathcal{N}=2$ SYK model obtained by deforming free fields by an SYK-like interaction was discussed in \cite{Bulycheva:2018qcp}.} The main reason this model is interesting is that one can show that it has a conformal manifold even at finite $N$. We will show this by taking a limit of the minimal models in section \ref{sec:GFF_limit_of_MM}.

The SUSY GFFs we will study is defined similarly to a standard (non-SUSY) GFF. We consider a complex superfield $\Phi$ whose two-point function takes the form
\begin{equation}\label{eq:GFF_prop}
    G(x_1,x_2)=\frac{1}{|\langle 12 \rangle|^{2\Delta}}\;,
\end{equation}
and so we identify it as having dimension $\Delta$
(see appendix \ref{app:conventions} for our $1+1$d $\mathcal{N}=2$ SUSY conventions). In addition, we take any $n$-point function of $\Phi$'s to reduce to the product of two-point functions obtained using all possible Wick contractions. In this way we have defined a CFT.

Now we can take $N$ such GFFs $\Phi_i$, and deform them by an SYK-like superpotential
\begin{equation}\label{eq:GFF_SYK}
W=\sum_{i_1...i_q} J_{i_1...i_q}\Phi_{i_1}...\Phi_{i_q}\;.
\end{equation}
The $J_{i_1...i_q}$'s are random couplings with variance proportional to $J^2$. We will call this model the disordered $\N=2$ GFFs.

We claim that we have a line of fixed points parametrized by $J$, in a similar manner to the cSYK model from the previous subsection. However, as we will show in \ref{sec:GFF_limit_of_MM}, the conformal manifold exists even for finite $N$, and so this model is a better example of conformal manifolds resulting from disorder. 

We can thus repeat the analysis of section \ref{sec:cSYK} in order to find the two- and four-point functions of the disordered GFFs.
First we solve the SD equation. As a reminder, for free chiral multiplets (GFFs with $\Delta=0$), this model is just the $1+1$d $\mathcal{N}=2$ SYK model, and the SD equations read \cite{Bulycheva:2018qcp}
\begin{equation}
D_1  \bar{D}_1  G(1  3)+J^2 \int d^2 z_{2} d^2 \theta_{2} G(1  2) G^{q-1}(3  2)=\left(\tilde{\theta}_1 -\tilde{\theta}_3 \right)\left(\overline{\tilde{\theta}}_1 -\overline{\tilde{\theta}}_3 \right) \delta(\langle 13\rangle) \delta(\langle\overline{1} \overline{3}\rangle)\;.
\end{equation}
In this case, $J$ is not an exactly marginal operator, and this equation is solved by neglecting the kinetic term in the IR and solving the rest of the equation using a conformal ansatz. Since our theory is conformal for all $J$, we should be able to solve the analogous equation without neglecting the kinetic term. 
For GFFs of dimension $\Delta$, the only change we must make is to change the kinetic term to 
\begin{equation}|D_1 |^{2-2\Delta} G(13)\;,
\end{equation}
where $\Delta=1/q$. Here, $|D_1 |^{2-2\Delta}$ denotes the inverse of the GFF propagator \eqref{eq:GFF_prop}.

As usual, we guess a scale-invariant solution of the form
\begin{equation}\label{eq:GFF_2_point}
G(12)=\frac{b(J)}{(\langle 12 \rangle\langle \bar 1 \bar 2 \rangle)^{\Delta}}
\end{equation}
for some constant $b(J)$. Plugging this into the equation, we find that (by definition) the kinetic term contributes a delta function, which we can move to the RHS. The equation becomes
\begin{equation}
J^2 \int d^2 z_{2} d^2 \theta_{2} G(1  2) G^{q-1}(3  2)=
(1-b)\left(\tilde{\theta}_1 -\tilde{\theta}_3 \right)\left(\overline{\tilde{\theta}}_1 -\overline{\tilde{\theta}}_3 \right) \delta(\langle 13\rangle) \delta(\langle\overline{1} \overline{3}\rangle)\;.
\end{equation}

We now proceed as in \cite{Bulycheva:2018qcp}. We find that for any $J$, there is a solution of the form \eqref{eq:GFF_2_point} where $b$ is a solution to the equation 
\begin{equation}
\frac{b^q}{1-b}=\frac{1}{4\pi^2 J^2}\;.
\end{equation}
As a consistency check, at $J=0$ we find $b=1$, while at $J\to\infty$ we reproduce the solution from \cite{Bulycheva:2018qcp}, see figure \ref{fig:GFFs_RG}. 
We have thus found that the exact two-point function for the disordered GFFs at any $J$ is identical to the one for the usual free fields, up to an overall factor of $\frac{b(J)}{b(J=\infty)}$ (as was the case for the cSYK model in section \ref{sec:cSYK}).

We can now compute the four-point function and the OTOC. Following the same logic as above, it is immediate to see that the kernel which appears in the four-point function is also given by the  kernel of the $1+1$d $\mathcal{N}=2$ SYK model, up to an overall factor of $\left(\frac{b(J)}{b(J=\infty)}\right)^q$. Explicitly, the eigenvalues of the bosonic part of the kernel (i.e.~the kernel for a four-point function whose external operators are all bottom components of our superfields) are given by
\begin{equation}
k^{GFF}(h,\tilde h)=\left(\frac{b(J)}{b(J=\infty)}\right)^qk^{\mathcal{N}=2\;SYK}(h,\tilde h)\;,
\end{equation}
where the eigenvalues for the bosonic part of the kernel for $1+1$d $\mathcal{N}=2$ SYK were found in \cite{Bulycheva:2018qcp}:
\begin{equation}
    k^{\mathcal{N}=2\;SYK}(h,\tilde h)=\Delta(1-\Delta) \frac{\Gamma^2(-\Delta)}{\Gamma^2(\Delta)} \frac{\Gamma(-h+\Delta) \Gamma(\tilde{h}+\Delta)}{\Gamma(1-h-\Delta) \Gamma(1+\tilde{h}-\Delta)}\;.
\end{equation}

\subsection{Disordered $\mathcal{N}=(2,2)$ minimal models}\label{sec:N2_MM}

\subsubsection{General $q$}

In this section we introduce a $1+1$d $\mathcal{N}=(2,2)$ conformal manifold. We will start with a UV description of this manifold in terms of a deformation of $N$ free fields, but ultimately we will have to study this model in conformal perturbation theory around a specific point on the manifold itself where the theory consists of $N$ decoupled minimal models. We will be interested in computing the chaos exponent as we vary the theory away from this special point, and specifically we will be interested in the way in which chaos appears.

\paragraph{UV description}~\\

The UV description of our theory includes $N$ chiral superfields $\Phi_i$ and a superpotential
\begin{equation}\label{eq:UV superpotential}
W=g\sum_{i=1}^N\Phi_{i}^q+\sum_{i_1,..., i_q} J_{i_1...i_q}\Phi_{i_1}...\Phi_{i_q}\;,
\end{equation}
where for now we focus on fixed realizations of the couplings $J_{i_1...i_q}$, and we are not averaging over them. 
Using the Leigh-Strassler argument \cite{Leigh:1995ep} one can show that there is a conformal manifold $\mathcal{M}_c$ in the space of couplings. We can count the number of exactly marginal operators by using the generating function of the chiral ring \cite{LERCHE1989427} (see \cite{Komargodski:2020ved} for a recent discussion). 
The generating function of the chiral ring for this theory is
\begin{equation}
P(t)=\Tr t^{J_R}=P(t)=\left( \frac{1-t^{2-2/q}}{1-t^{2/q}} \right)^{N}\;.
\end{equation}
Here, $J_R$ is the generator of the R-symmetry, and the trace is over all chiral primaries. Explicitly, the coefficient of $t^r$ is the number of chiral primaries with R-charge $r$. In particular, the coefficient of $t^2$ is $\dim\mathcal{M}_c$, or the number of exactly marginal operators in the theory (since chiral operators have dimension $\Delta=\frac{r}{2}$ in $1+1$d $\mathcal{N}=2$ SCFTs).
Then in the large-$N$ limit
\begin{equation}
\dim \mathcal{M}_c=\frac{1}{q!}N^{q}+O(N^{q-1})\;.
\end{equation}
We thus find a huge conformal manifold in the large-$N$ limit of this theory. In fact, a direct computation shows that in the large-$N$ limit, ``most'' classically marginal operators become exactly marginal, in the sense that
\begin{equation}\label{eq:IRR_vs_EM}
\frac{\# CM-\# EM}{\# CM}=\frac{q!}{N^{q-2}}+O(1/N^{q-1}) \;,
\end{equation}
where $\# CM$ denotes the number of classically marginal operators and $\# EM=\dim \mathcal{M}_c$ denotes the number of exactly marginal operators.

We thus expect the theory to have a large conformal manifold at large $N$. However, from the UV point of view the conformal manifold is strongly-coupled, and so the Lagrangian \eqref{eq:UV superpotential} is not useful for computations. Instead, we next consider conformal perturbation theory around a specific point on this conformal manifold.

\paragraph{IR description}~\\

Next, we move on to the IR description. This involves two steps. First, we flow to the IR CFT defined by setting $J_{i_1...i_q}=0$ in \eqref{eq:UV superpotential}. This describes $N$ decoupled copies of a CFT defined by the superpotential 
\begin{equation}
W=\Phi^q\;.
\end{equation}
This theory is known to flow to the $\mathcal{N}=(2,2)$ $A_{q-1}$ minimal model \cite{Witten:1993jg}. It includes a chiral multiplet $\Phi^{IR}$ with dimension $\Delta=1/q$, and its central charge is $c=3(1-2/q)$ \cite{LERCHE1989427}. It has no continuous global (non-R) symmetries.

Next, we deform the $N$ decoupled minimal models:
\begin{equation}\label{eq:IR_model}
W=\sum_{i=1}^N\Phi_{i}^q+\sum_{i_1\neq....\neq i_q} J^{IR}_{i_1...i_q}\Phi^{IR}_{i_1}...\Phi^{IR}_{i_q}\;.
\end{equation}
We emphasize the interpretation of this superpotential; we have assumed that we first tune the first term to the CFT, and then deform by the second term. We will call this model the disordered $\N=2$ $A_{q-1}$ minimal model.

We can now argue that each deformation $J^{IR}_{i_1...i_q}$ is exactly marginal. First, since $\Delta_{\Phi_i^{IR}}=1/q$, the deformation is classically marginal. However, since the CFT at $J^{IR}=0$ has no continuous non-R global symmetries, every classically marginal operator is exactly marginal \cite{Green:2010da,Kol:2002zt,Kol:2010ub}.\footnote{It is crucial for this argument that we sum over $i_1\neq...\neq i_q$ in \eqref{eq:IR_model} in order for every deformation to be a nonzero element of the chiral ring.} We thus learn that every realization of the model is conformal. Clearly we end up on the same conformal manifold as described in the UV description above.

\paragraph{Summary}~\\

We have thus found that the disordered $\N=2$ minimal models defined in \eqref{eq:IR_model} describe a conformal manifold. This model should be understood as deforming $N$ copies of the $\mathcal{N}=(2,2)$ $A_{q-1}$ minimal model by a disorder superpotential. This conformal manifold is the same as the one expected to appear had we started in the UV from free fields and deformed by a similar superpotential. 

We emphasize that the disorder average is an average over CFTs, i.e.~we have a CFT at every realization of the theory, as opposed to the disorder average in the standard SYK model. This is similar to other recent examples where averages were performed over CFTs, see e.g.~\cite{Maloney:2020nni,Afkhami-Jeddi:2020ezh}.

One can now perform computations in this model. We will be interested in the two- and four-point functions of $\Phi_i^{IR}$. In the following, we will focus on the IR description, and we will suppress the ``IR'' indices for clarity. These computations require performing conformal perturbation theory around the CFT at $J=0$, which is $N$ copies of the $A_{q-1}$ minimal model. This is difficult in general, and so in the following we will focus on the particularly simple example of this model with $q=3$, which we discuss next.

\subsubsection{The case $q=3$}\label{sec:q3_Case}

In this section we focus on the case $q=3$ of the conformal manifold defined in \eqref{eq:IR_model}, which is a specifically simple case. Explicitly, the superpotential is 
\begin{equation}
W=\sum_{i=1}^N\Phi_{i}^3+\sum_{i\neq j\neq k} J_{ijk}\Phi_{i}\Phi_{j}\Phi_{k}\;,
\end{equation}
where we are still interpreting the theory as a deformation of $N$ copies of the $A_{2}$ minimal models at $J=0$, but we have suppressed the ``IR'' index in the second term. In particular, in the following $\Phi$ should be interpreted as a chiral superfield in the $A_{2}$ minimal model with dimension $\Delta=1/3$.

We start by analyzing the model at $J=0$. Consider a single copy of the $A_2$ $\mathcal{N}=2$ minimal model, which has a chiral superfield $\Phi$ of dimension $1/3$ and central charge $c=1$. As a result, this CFT should be dual to the free compact boson at some special value of the radius $R$. We discuss the details of this duality in appendix \ref{app:q3_min_model}. We can thus think of the model at $J=0$ as being $N$ copies of the $c=1$ compact boson at a specific value of the radius $R$.

In order to proceed, we must match the components of the superfield $\Phi$ with vertex operators in the $c=1$ theory. This is done in appendix \ref{app:q3_min_model}. Since we know the exact form of any $n$-point function of vertex operators, we now immediately have all $n$-point functions of the $\Phi_i$'s at $J=0$. For our purposes, it would be most useful to have these in superspace, instead of in components. In appendix \ref{app:q3_min_model} we present a conjecture for the form of these $n$-point functions in superspace (see our superspace conventions in appendix \ref{app:conventions}), which has been checked explicitly for the case of the 2,4 and 6-point functions. This form is
\begin{equation}
\langle \Phi(x_1)...\Phi(x_n)\bar \Phi(y_1)...\bar \Phi(y_n)\rangle=\left|\sum_{\sigma\in S_n} \text{sign}(\sigma)\prod_{i=1}^n \frac{1}{\langle x_i-y_{\sigma(i)}\rangle}\right|^{2\Delta}\;,
\end{equation}
where all $\Phi$'s are taken from the same copy of the minimal model (otherwise the correlator at $J=0$ decouples). 
At leading order in $J$ we will only need the 4-point function, which is explicitly
\begin{equation}\label{eq:4_X}
\langle \bar \Phi \Phi \bar \Phi \Phi \rangle =\left|\frac{1}{\langle 12\rangle\langle 34\rangle}\right|^{2\Delta}|1-\chi_S|^{2\Delta}\;,
\end{equation}
where
\begin{equation}\label{eq:chiS_main_text}
\chi_S=\frac{\langle 12\rangle \langle 34\rangle }{ \langle 14\rangle \langle 32\rangle }\;,\qquad
\langle 12\rangle = z_{12}-2\tilde \theta_1\theta_2-\theta_1\tilde\theta_1 -\theta_2\tilde\theta_2\;,
\end{equation}
and with $\Delta=1/3$. 

As a result, the subtracted four-point function $4_s$ defined in section \ref{sec:2_pt_SD} takes the form
\begin{equation}\label{eq:4_X_subtracted}
\langle \bar \Phi_i \Phi_i \bar \Phi_i \Phi_i \rangle_s =\left|\frac{1-\chi_S}{\langle 12\rangle\langle 34\rangle}\right|^{2\Delta}-
\left|\frac{1}{\langle 12\rangle\langle 34\rangle}\right|^{2\Delta}\;.
\end{equation}
We can now use these results to study the disordered theory using the analysis of section \ref{sec:chaos_nontrivial_CFT} (in particular, we interpret the diagrams in section \ref{sec:chaos_nontrivial_CFT} as supergraphs). This will allow us to study chaos for the $q=3$ disordered SUSY minimal models in section \ref{sec:min_models_chaos}.

\subsection{Disordered GFFs as a limit of the $N_f$-flavored minimal models}\label{sec:GFF_limit_of_MM}

Above we described two models: the disordered $\mathcal{N}=2$ GFFs in section \ref{sec:2d_GFFs} and the disordered $\mathcal{N}=2$ minimal models in section \ref{sec:N2_MM}. We now discuss a generalization of the disordered $\mathcal{N}=2$ minimal models, from which the GFFs can be obtained using a specific limit. This will allow us to show that there is a conformal manifold in the disordered SUSY GFF theory for $\Delta=1/q$ even at finite $N$ and for every realization of $J_{i_1...i_q}$.

We build the model in a similar fashion to the disordered $\N=2$ minimal models from section \ref{sec:N2_MM}. We start with $N\times N_f$ chiral superfields $\Phi_{i,a}$, with $i=1,...,N$ and $a=1,...,N_f$. Adding a superpotential 
\begin{equation}
W=\sum_{a=1}^{N_f}\sum_{i=1}^N\Phi_{i,a}^q\;,
\end{equation}
we can flow to $N\times N_f$ copies of the $A_{q-1}$ minimal model. Next, add an SYK-like interaction:
\begin{equation}\label{eq:Nf_flavored_min_models_IR}
W=\sum_{a=1}^{N_f}\sum_{i=1}^N\Phi_{i,a}^q+\sum_{a_1,...,a_q}\sum_{i_1\neq ...\neq i_q} \tilde{J}_{i_1...i_q}\Phi^{IR}_{i_1,a_1}...\Phi^{IR}_{i_q,a_q}\;.
\end{equation}
We can repeat the arguments in section \ref{sec:N2_MM} to learn that we have a conformal manifold for each realization of the $\tilde{J}_{i_1...i_q}$'s. We think of this model as having $N_f$ ``flavors'' of the original $\N=2$ disordered minimal models (a somewhat similar construction for SYK was discussed in \cite{Gross:2016kjj}).

We can thus average over the $\tilde{J}_{i_1...i_q}$ assuming they are again random variables with a Gaussian distribution, this time with variance $\langle \tilde{J}_{i_1...i_q}^2\rangle=(q-1)!\frac{\tilde{J}^2}{N^{q-1}N_f^q}$ (note the dependence on $N_f$). As explained above, since the dimension of $\Phi^{IR}$ is fixed by superconformal invariance, we expect its correlators to have the conformal form even after averaging.

We will now show that in the limit $N_f\to\infty$, the theory reduces to that of the disordered GFFs (this is not surprising - indeed, it is similar to how one can obtain GFFs in higher dimensions by taking the large-$N$ limit of certain gauge theories). Define the operator
\begin{equation}
\Psi_i=\frac{1}{\sqrt{N_f}}\sum_a \Phi_{i,a}\;.
\end{equation}
Then the disorder interaction term in the superpotential above can be written as
\begin{equation}
\sum_{i_1...i_q}J_{i_1...i_q}\Psi_{i_1}...\Psi_{i_q}\;.
\end{equation}
The normalization of $J_{i_1...i_q}$ is now the same as for the standard disorder deformations discussed above, $\langle J_{i_1...i_q}^2\rangle=(q-1)!\frac{J^2}{N^{q-1}}$.
Then it is clear that we can compute the exact two-point and four-point functions of $\Psi$ by using the methods in section \ref{sec:chaos_nontrivial_CFT}. In particular, this requires knowing the exact $n$-point functions of $\Psi_i$ at the CFT at $J=0$. But at leading order in $1/N_f$, these are particularly simple, and they reduce to products of 2-point functions.

For example, consider the 4-point function. Suppressing positions of operators, this can be written as
\begin{equation}
\langle \Psi_i\overline\Psi_i\Psi_i\overline\Psi_i \rangle = \frac{1}{N_f^2}\sum_{a,b,c,d} \langle \Phi_{ia}\overline\Phi_{ib}\Phi_{ic}\overline\Phi_{id} \rangle
= \frac{1}{N_f^2}\sum_{a,b,c,d} (\delta_{ab}\delta_{cd}+\delta_{ad}\delta_{bc}) \langle \Phi_{ia}\overline\Phi_{ia} \rangle^2 +\delta_{abcd}\langle \Phi_{ia}\Phi_{ia}\Phi_{ia}\Phi_{ia} \rangle\;.
\end{equation}
Performing the sums, we find
\begin{equation}
\langle \Psi_i\overline\Psi_i\Psi_i\overline\Psi_i \rangle 
=  \langle \Phi_{ia}\overline\Phi_{ia} \rangle^2 +\frac{1}{N_f}\langle \Phi_{ia}\Phi_{ia}\Phi_{ia}\Phi_{ia} \rangle\;.
\end{equation}
where there is no sum over repeated indices. It is then clear that the leading contribution to the four-point function comes from the disconnected diagrams which connect $\Psi$'s using propagators, and that the connected four-point function only contributes at subleading order in $1/N_f$. Similar arguments can be used to show that higher $n$-point functions also reduce to products of propagators at leading order in $1/N_f$. 

We have thus found that at leading order in $1/N_f$, this model behaves as if it were a theory of GFFs with dimension $\Delta=1/q$. We can thus find its two- and four-point functions using the results of \ref{sec:N2_MM}. In particular, this proves that there is a conformal manifold in the disordered GFF theory (unlike in the cSYK model, for which there is evidence of a conformal manifold only at leading order in $1/N$).

\section{Chaos in disordered generalized free fields} \label{sec:chaos_GFF}

\subsection{The cSYK model}\label{sec:cSYK_chaos}

In section \ref{sec:cSYK}, we discussed the cSYK model, in which $J^2$ was an exactly marginal operator at leading order in $1/N$. We can now compute the chaos exponent in this theory as a function of $J$. 

As discussed in \ref{sec:cSYK}, the two-point function of cSYK is identical to that of SYK, up to an overall function $\bar b(J)$. As a result, the kernel of cSYK is also proportional to that of SYK, up to the proportionality factor \eqref{eq:K_cSYK}. One can repeat the same argument also for the retarded kernel. The result is that the retarded kernel $K_R$ for cSYK is identical to the retarded kernel for SYK up to the same overall factor $\left(\frac{\bar{b}(J)}{\bar b(J\to \infty)}\right)^{q} $. Using the result for the eigenvalues in SYK \eqref{eq:SYK_kR_eigenvalues}, we find
\begin{equation}\label{eq:GFFs_kR}
k_R(\lambda)=\left(\frac{\bar{b}(J)}{\bar b(J\to \infty)}\right)^{q}\frac{\Gamma(3-2 \Delta) \Gamma(2 \Delta+\lambda)}{\Gamma(1+2 \Delta) \Gamma(2-2 \Delta+\lambda)}\;.
\end{equation}
We can now find the chaos exponent by looking for the largest value of $\lambda_L$ such that $k_R(\lambda_L)=1$. 

\begin{figure}
	\centering
	\begin{subfigure}[t]{0.5\textwidth}
		\centering
		\includegraphics[width=0.8\linewidth]{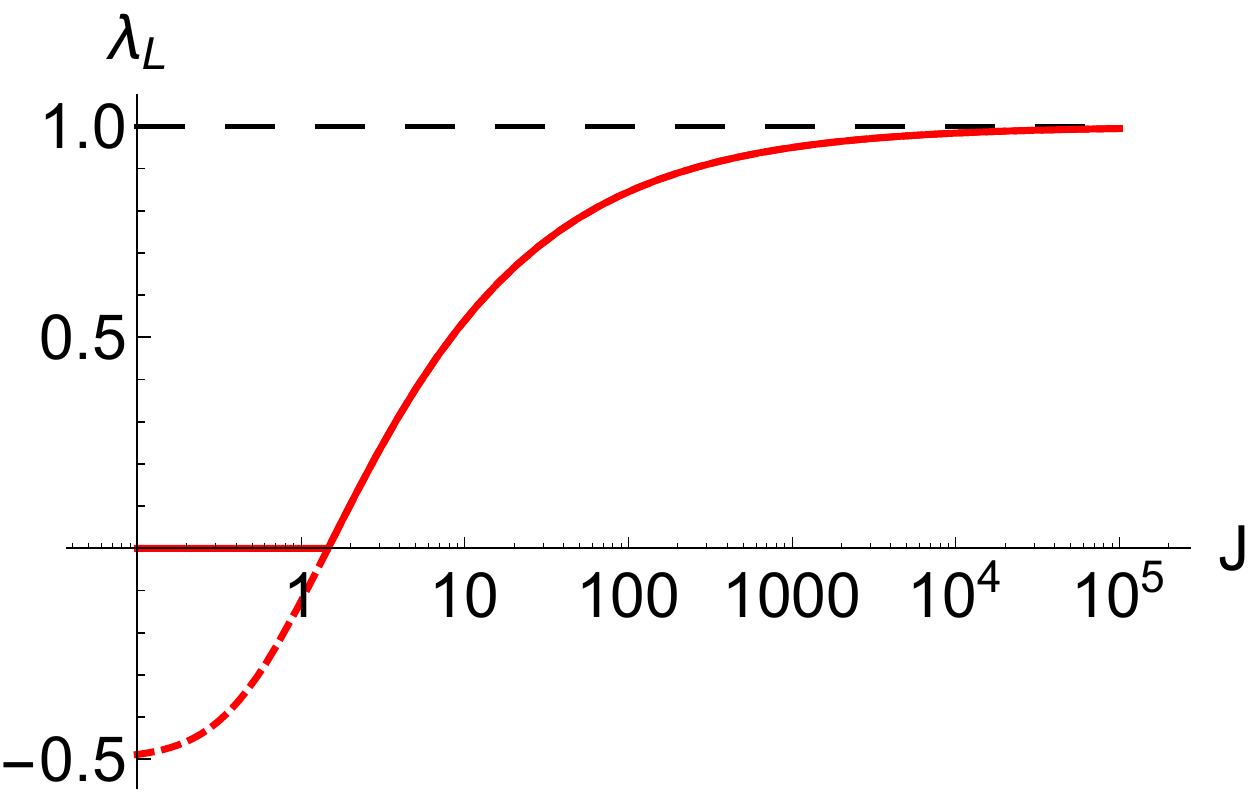}
		\caption{}
		\label{fig:GRchaosa}
	\end{subfigure}%
	~ 
	\begin{subfigure}[t]{0.5\textwidth}
		\centering
		\includegraphics[width=1\linewidth]{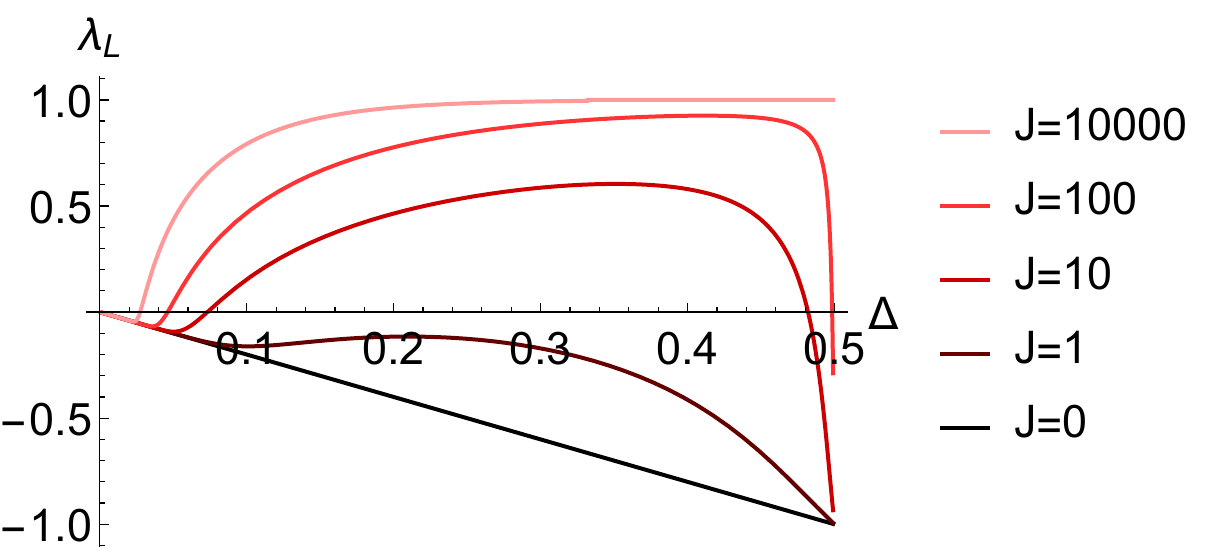}
		\caption{}
		\label{fig:GRchaosb}
	\end{subfigure}
	\caption{The chaos exponent for QM GFFs. (a) $\ly$ as a function of $J$ at $\Delta=0.25$. The dashed red line represents the solution to $k(\lambda_L)=1$, but wherever the solution is negative we interpret $\lambda_L$ to be zero there, corresponding to the solid line. At large $J$ the solutions asymptotes to $\ly=1$. (b) $\ly$ as a function of $\Delta$ for various values of $J$. The black line is $\lambda_L=-2\Delta$ and corresponds to $J=0$. This is negative for $\Delta>0$, leading to the discontinuous transition into chaos.}
	\label{fig:GRchaos}
\end{figure}

We start by plotting the result for the specific value $q=4$ in figure \ref{fig:GRchaosa}. At large $J$ the result goes to the maximal value $\ly=1$, as expected. The behavior at small $J$ is more interesting. At small enough $J$, the solution to $k_R=1$ becomes negative for $\Delta>0$, corresponding to the dashed red line (in particular, at $J=0$ we find $\lambda_L(J=0)=-2\Delta$). This is unphysical, as discussed in section \ref{sec:chaos_summary}; in computing $\ly$ we assumed it was positive, and so this solution is not self-consistent. However, we immediately learn that there cannot be any solution with positive $\ly$ in this region, since our calculation would have found it. We thus conclude that there is no chaos in this region, meaning that the correct value is $\ly=0$, corresponding to the solid red line. We thus find a discontinuous transition into chaos, reminiscent of KAM theory.

Finally, we plot the general form for the chaos exponent for any $q$ in Figure \ref{fig:GRchaosb}. Note that at large enough $J$, the chaos exponent always asymptotes to $\lambda_L=1$ regardless of $q$, as expected. More interestingly, note that at small $J$, $\lambda_L$  asymptotes to $-2\Delta=-\frac{2}{q}$. As discussed in section \ref{sec:chaos_summary}, a negative value for $\lambda_L$ should be interpreted as having zero chaos in this region, $\lambda_L=0$, corresponding to the solid line.  In particular, for any $q<\infty$, we will find a discontinuous, KAM-like, behavior. This is inline with the discussion of section \ref{sec:chaos_J0}: taking $J \to 0$ at fixed $\lambda$, the eigenvalues \eqref{eq:GFFs_kR} vanishes, and so in order to find solutions to the eigenvalue equation $k_R=1$ we must look for values of $\lambda$ for which $k_R$ diverges at fixed small $J$. Indeed, one can check that $k_R$ diverges as $\lambda=-2\Delta$. 

As we will see in section \ref{sec:validity_1}, for this theory is also $\lambda_L^0=-2\Delta$. In other words, in this case the continuity conjecture \eqref{eq:continuity} holds.

\subsection{The disordered $1+1$d $\mathcal{N}=2$ GFFs}

We now compute the chaos exponent for the disordered GFFs discussed in section \ref{sec:2d_GFFs}. This conformal manifold is obtained by taking $N$ decoupled $1+1$d $\mathcal{N}=2$ GFFs, and adding an SYK-like superpotential. The model is very similar to the cSYK model described above, and we will one again be able to compute the chaos exponent as a function of the deformation parameter $J$.

\begin{figure}[t]
	\centering
	\begin{subfigure}[t]{0.5\textwidth}
		\centering
		\includegraphics[width=0.8\linewidth]{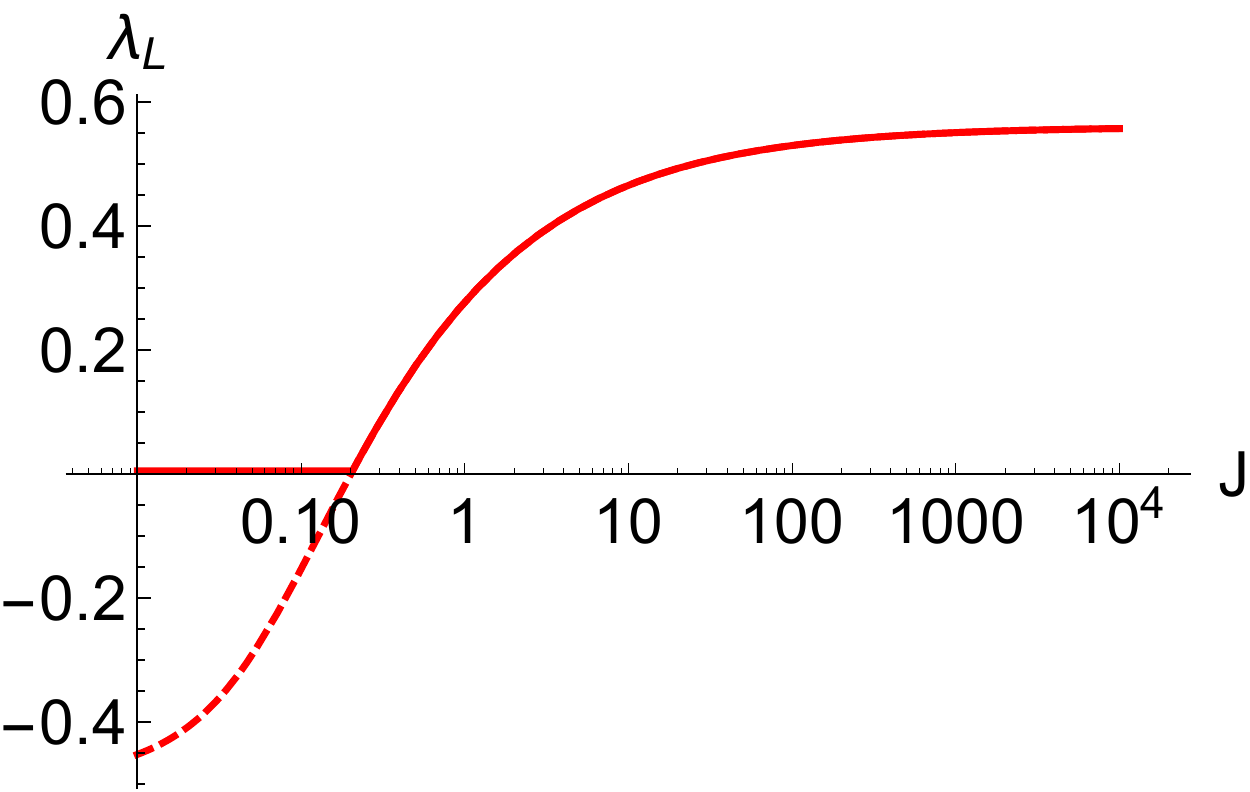}
		\caption{}
	    \label{fig:2dGFFchaosa}
	\end{subfigure}%
	~ 
	\begin{subfigure}[t]{0.5\textwidth}
		\centering
		\includegraphics[width=1\linewidth]{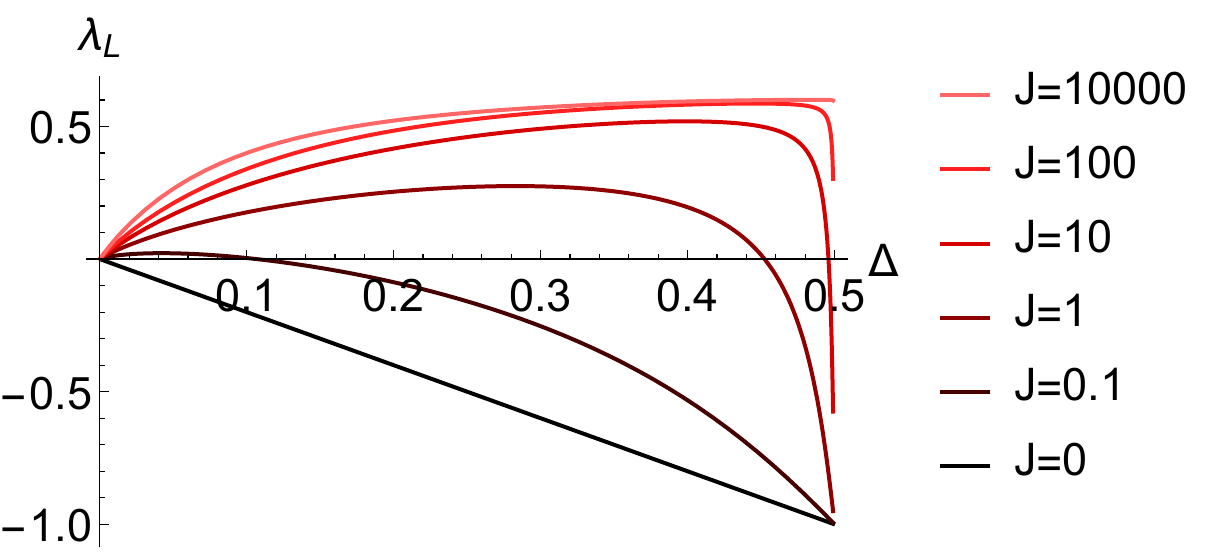}
		\caption{}
		
	\end{subfigure}
	\caption{The chaos exponent for $1+1$d SUSY GFFs. (a) The chaos exponent as a function of $J$ at $\Delta=0.25$. The dashed line represents the solution to $k(\lambda_L)=1$, but wherever the solution is negative we interpret $\lambda_L$ to be zero there, corresponding to the solid line. (b)  The chaos exponent as a function of $\Delta$ for various values of $J$. Again, for small enough $J$, $\lambda_L$ is negative.}
	\label{fig:2dgffchaos}
\end{figure}

As discussed in \ref{sec:2d_GFFs}, the exact two-point function for the GFFs is identical to the one for the usual $\mathcal{N}=2$ SYK model obtained by deforming free fields, up to an overall factor of $\frac{b(J)}{b(J=\infty)}$. Thus, following the same logic as in the previous subsection, the retarded kernel is also given by the retarded kernel of the $1+1$d $\mathcal{N}=2$ SYK model, up to an overall factor. Explicitly, the eigenvalues of the bosonic part of the retarded kernel are given by
\begin{equation}
k_R^{GFF}(h,\tilde h)=\frac{b(J)}{b(J=\infty)}k_R^{\mathcal{N}=2\;SYK}\;,
\end{equation}
where the eigenvalues for the $1+1$d $\mathcal{N}=2$ SYK were found in \cite{Bulycheva:2018qcp}:
\begin{equation}
k_R^{\mathcal{N}=2\;SYK}=-\frac{\Gamma^2(1-\Delta)}{\Gamma(\Delta+1)\Gamma(\Delta-1)}\frac{\Gamma(\Delta-h)\Gamma(\Delta-\tilde h)}{\Gamma(1-h-\Delta)\Gamma(1-\tilde h-\Delta)}\;.
\end{equation}
The chaos exponent is then found by finding solutions to $k_R(h,\tilde h)=1$. The result for the chaos exponent appears in figure \ref{fig:2dgffchaos}.
We have also checked explicitly that other components of the kernel other than the bosonic one don't contribute a larger chaos exponent.

Again, there are a couple of interesting features to notice. First, at large enough $J$, the chaos exponent for any $\Delta$ approaches the result in the $1+1$d SUSY version of the SYK model discussed in \cite{Murugan:2017eto,Bulycheva:2018qcp}, as expected. Second, for any $\Delta>0$, for small enough $J$ the chaos exponent becomes negative. As a result, we again find a discontinuous transition into chaos for any $\Delta>0$. This can be seen explicitly in figure \ref{fig:2dGFFchaosa}, where we see that indeed the solution to $k(\lambda_L)=1$ becomes negative below some critical $J$ (corresponding to the dashed line in the figure). We have thus found that the disordered $1+1$d SUSY GFFs also display a discontinuous transition into chaos, similar to the $0+1$d case discussed above.

\subsection{Checking the validity of the approximation}\label{sec:validity_1}

In defining the chaos exponent, we had to solve equation \eqref{eq:4pt_Self_consistency}:
\begin{equation}
    W_R=F_0+K_RW_R\;.
\end{equation}
We assumed that $W_R$ grows exponentially, which allowed us to neglect the $F_0$ term. Now we return to this assumption and check it explicitly. We do this for the $0+1$d GFFs, and the result for the $1+1$d version is similar. $F_0$ is given by a product of propagators:
\begin{equation}
    F_0=G_R(13)G_R(24)\;,
\end{equation}
see figure \ref{fig:freeKer}. Using the explicit form of $G_R$ for SYK (see e.g. \cite{Murugan:2017eto})
\begin{equation}
    G_{R}\left(t, t^{\prime}\right)=\theta\left(t-t^{\prime}\right) \frac{2 b \cos \pi \Delta}{\left(2 \sinh \frac{1}{2}\left(t-t^{\prime}\right)\right)^{2 \Delta}}\;,
\end{equation}
and using the fact that for GFFs $G_R$ is proportional to the one for SYK, we find at large $t_3,t_4$ that
\begin{equation}
    F_0\propto e^{\Delta(t_3+t_4)}\;.
\end{equation}
Comparing to equation \eqref{eq:exponential_growth}, we can read off the chaos exponent predicted by $F_0$ at $J=0$: \begin{equation}
    \lambda_L^{0}=-2\Delta\;.
\end{equation}
However, we found above that neglecting $F_0$, the chaos exponent $\lambda_L(J)$ is at least $\lambda_L(J=0)=-2\Delta$, and so neglecting $F_0$ above is justified.

In particular, we have found that the chaos exponent predicted by $F_0$ matches with the limit of the chaos exponent predicted by the kernel as $J\to 0$. This means that  we have verified the continuity relation $\lambda_L^{0}=\lambda_L(J\to 0)$ (see equation \eqref{eq:continuity}) for these two examples.

\section{Chaos in the disordered $\mathcal{N}=2$ $A_2$ minimal models}\label{sec:min_models_chaos}

In this section we compute the chaos exponent for the disordered SUSY minimal models with $q=3$. While the calculation of the chaos exponent for the disordered GFFs was possible for all $J$, it is much more complicated for the disordered minimal models due to their nontrivial $n$-point functions. We will thus only be able to compute the chaos exponent at small $J$. In addition, we will only be able to do this for the particularly simple case $q=3$, where we know all $n$-point functions of the CFT, see section \ref{sec:q3_Case}.

\subsection{Computing $\lambda_L$ near $J=0$}

We now compute the chaos exponent for small $J$ for the disordered $\mathcal{N}=2$ $A_2$ minimal model. As discussed in section \ref{sec:chaos}, at leading order in $J$ this requires knowing only the full 4-point function of the undeformed CFT, given in equation \eqref{eq:4_X}. We can then plug this into (the supersymmetric version of) equation \eqref{eq:kR_def} and compute the eigenvalues of the kernel $K_R$ at leading order in $J$. 

We again focus on the bosonic eigenvalue $k_R$.\footnote{We will assume that the bosonic eigenvalue will give the leading chaos exponent. Indeed this was the case in all previous examples considered.} As discussed in section \ref{sec:chaos}, $k_R$ is given by computing
\begin{equation}\label{eq:eigenvalues_MM}
     k_R(\lambda,J)=\frac{K_R\cdot W}{W}=\frac{\int d^2x_3 d^2 \theta_3 d^2 x_4 d^2\tilde\theta_4K_R(x_1,x_2;x_3,x_4) W(x_3,x_4) }{W(x_1,x_2)}\;,
\end{equation}
where the eigenfunction $W$ is the superspace generalization of \eqref{eq:W_ansatz}:
\begin{equation} \label{eq:susy_W}
\begin{split}
    W(1,2) &= \frac{\exp(-\frac{h+\tilde h}{2}(t_1+t_2)+\frac{h-\tilde h}{2}(x_1+x_2))}
    {(2\cosh(\frac{t_{12}-x_{12}}{2})-i \tilde\theta_1 \theta_2)^{\Delta-h}
    (2\cosh(\frac{t_{12}+x_{12}}{2})+i \bar\tilde\theta_1 \bar\theta_2)^{\Delta-\tilde h}}\;.
\end{split}
\end{equation}
We can change variables to 
\begin{equation}
    \begin{split}
        u_3 = e^{x_3-t_3}\;,&\quad v_3 = e^{-x_3-t_3}\;, \\ \theta_3^u = e^\frac{x_3-t_3}{2} \theta_3\;,&\quad \theta_3^v = e^\frac{-x_3-t_3}{2} \bar\theta_3\;,\\
        \tilde\theta_3^u = e^\frac{x_3-t_3}{2} \tilde\theta_3\;, &\quad \tilde\theta_3^v = e^\frac{-x_3-t_3}{2} \bar{\tilde{\theta}}_3\;,
    \end{split}
\end{equation} 
and similarly 
\begin{equation}
    \begin{split}
    u_4 = -e^{x_4-t_4}\;, & \quad v_4 = -e^{-x_4-t_4}\;,\\ \theta_4^u = i e^\frac{x_4-t_4}{2} \theta_4\;, & \quad \theta_4^v = i e^\frac{-x_4-t_4}{2} \bar\theta_4\;,\\ \tilde\theta_4^u = -i e^\frac{x_4-t_4}{2} \tilde\theta_4\;, & \quad \tilde\theta_4^v = e^\frac{-x_4-t_4}{2} \bar{\tilde\theta}_4\;, 
    \end{split}
\end{equation}  and find that at leading order in $J$ this is given by
\begin{equation}
    k_{R}|_{J^2}= \frac{J^2}{2} \frac{1}{W(1,2)\mid_{\theta_{1,2}=0}}\int  \frac{du_3 dv_3 d\theta^u_3 d\theta^v_3}{\langle 34\rangle _u^{1-h}}
    \frac{du_4 dv_4 d\tilde\theta^u_4 d\tilde\theta^v_4}{\langle 34\rangle_v^{1-\tilde h}}
    \mathcal{G}_R(\chi_S,\bar\chi_S)\;.
\end{equation}
Here we have defined $\langle 34\rangle_u = u_{34} - 2  \theta_3^u \tilde\theta_4^u$ and $\langle 34\rangle_v = v_{34} - 2  \theta_3^v \tilde\theta_4^v$. $\mathcal{G}_R$ is the retarded normalized four-point \eqref{eq:def_G_R}, and $\chi_S, \bar\chi_S$ are given in \eqref{eq:susy_ratio}. 

Instead of finding the value of $\mathcal{G}_R$ directly, we will first do the Grassman integral without analytically continuing in $u,v$, and then analytically continue (as the two operations commute). Using \eqref{eq:4_X_subtracted}, the superspace integral gives
\begin{equation} 
\begin{split}
K\cdot & W|_{J^2}=\frac{J^2}{2} \frac{1}{W(1,2)\mid_{\theta_{1,2}=0}} 
\int \frac{du_3 dv_4 du_4 dv_4} {(u_{34})^{2-h}(v_{34})^{2-\tilde h}}\\
&\cdot\bigg(\left|1-\chi\right|^{2/3}\left(2(1-h)+\frac{2\Delta\chi}{1-\chi}\right)\left(2(1-\overline{h})+\frac{2\Delta\overline{\chi}}{1-\overline{\chi}}\right)
-4(1-h)(1-\overline{h})\bigg)_R\;,
\end{split}
\end{equation}
where by $(...)_R$ we mean the same operation taken in eq. \eqref{eq:def_G_R}, and  $\chi=\frac{u_{12}u_{34}}{u_{14}u_{23}},\bar\chi=\frac{v_{12}v_{34}}{v_{14}v_{23}}$. Focusing on the case $h=\overline{h}=-\frac{\lambda}{2}$ (as explained above), we can simplify this expression to 
\begin{equation} 
K\cdot W|_{J^2}=\frac{J^2}{2} \frac{1}{W(1,2)\mid_{\theta_{1,2}=0}}
\int \frac{du_3 dv_4 du_4 dv_4} {(u_{34} v_{34})^{1+\frac{\lambda}{2}}}
\left(\left|1-\chi\right|^{2/3}\left|2(1+\frac{\lambda}{2})+\frac{2\Delta\chi}{1-\chi}\right|^2-4
\left(1+\frac{\lambda}{2}\right)^2
\right)_R\;.
\end{equation}
Written in this way, the analytic continuation is straightforward,\footnote{The analytic continuation can be done almost immediately by showing that the integrand can be written in terms of products of propagators between the points $z_1,z_2,z_3,z_4$, and replacing these propagators with the relevant analytically continued propagators, denoted by $G_{lr}$ and $G_R$ in \cite{Murugan:2017eto}.} and we find
\begin{equation}\label{eq:4point_integral}
\begin{split}
    K_R\cdot W|_{J^2}=&\frac{J^2}{2} \frac{1}{W(1,2)\mid_{\theta_{1,2}=0}}
\int du_3 dv_4 du_4 dv_4
    \bigg(
    \left(1+\frac{\lambda}{2}\right)
    \left(\frac{2}{3}+\frac{\lambda}{2}\right)
    \frac{\left(\sin\frac{\pi}{3}\right)^2}{|34|^{4+\lambda}}
    \left(
    \frac{|13||24|}{|14||23|}
    \right)^{2/3}.\\
    &+\frac{1}{3}\left(\frac{\lambda}{2}+\frac{2}{3}\right)
    \frac{\left(\sin\frac{\pi}{3}\right)^2 |12|^2}
    {|34|^{2+\lambda}(|13||24|)^{4/3}(|14||23|)^{2/3}}
    +\frac{1}{3}
    \left(1+\frac{\lambda}{2}\right)\frac{\left(\sin\frac{2\pi}{3}\right)^2}{|34|^{4+\lambda}}
    \left(\frac{|14||23|}{|13||24|}\right)^{4/3} \bigg)\\
    &+(3\leftrightarrow4)\;.
\end{split}
\end{equation}
Here we denoted $|ij|^2=(u_i-u_j)(v_i-v_j)$ for brevity.
In particular, the contribution of the ``subtraction'' term proportional to $(1-h)^2$ vanishes. As a consistency check, we have checked numerically that using this expression, the eigenvalues \eqref{eq:eigenvalues_MM} are indeed independent of the external points $1,2$.

As discussed in section \ref{sec:chaos}, the value of $\lambda_L$ close to $J=0$ is found by looking for values of $\lambda$ for which the integral diverges in the limit $|u_i|,|v_i|\rightarrow \infty$. Dimensional analysis shows that the largest value of $\lambda$ for which the integral diverges is $\lambda=\lambda_L=0$, and a numerical computation of the integral confirms this. We thus find that the chaos exponent near $J=0$ is $\lambda_L(J\to 0)=0$. The transition into chaos will thus be continuous (assuming $\lambda_L$ grows with increasing $J$), as in figure \ref{fig:continuous_chaos}.

As discussed in section \ref{sec:chaos}, consistency of our perturbative expansion requires that we check that contributions from higher $n_s'$-point functions diverge at values of $\lambda$ which are at most $\lambda_L=0$.\footnote{It must also be checked that the subleading correction is positive, i.e.~that the chaos exponent rises as we raise $J$. We will not check this explicitly but instead assume this is the case. Indeed this has been the case in all previous examples considered.} We will check this order-by-order in $n$.
The $2n$-point function appears in \eqref{eq:n_point_conjecture}:
\begin{equation}\label{eq:n_point_conjecture_main_text}
\langle \Phi(x_1)...\Phi(x_n)\bar \Phi(y_1)...\bar \Phi(y_n)\rangle=\left|\sum_{\sigma\in S_n} \text{sign}(\sigma)\prod_{i=1}^n \frac{1}{\langle x_i-y_{\sigma(i)}\rangle}\right|^{2\Delta}\;.
\end{equation}
The $(2n)_s'$ correlator is obtained from this correlator using various subtractions of lower $n$-point functions with legs contracted using $\Sigma$'s. First we explain why we can ignore the subtractions when discussing the leading divergence at order $2n$. There are two types of subtractions in the four-point functions: subtractions of disconnected diagrams where the points $x_1,x_2$ are disconnected from $x_3,x_4$, and subtractions where $x_1,x_2$ are connected to $x_3,x_4$ but there are additional disconnected bubbles in the diagram. In the former case, the diagrams vanish in the limit $\epsilon\to 0$, and so we can ignore them. In the latter case, what we find is that if we ignore the bubble diagrams, the remaining part which connects $x_1,x_2$ to $x_3,x_4$ is identical to some lower-order $(2n)_s'$ correlator, and so the divergence from it already appears at lower orders in $n$ and we have taken it into account.

It it thus enough to plug in the $(2n)$-point function \eqref{eq:n_point_conjecture_main_text} into the diagram and compute at what values of $\lambda$ its contribution diverges. The integral that appear in $k_R$ at order $J^{2n+2}$ is
\begin{equation}
\begin{split}
    \prod_{i=1}^{n}\int\frac{d^{2}x_{i}d^{2}\theta_{x_{i}}d^{2}y_{i}d^{2}\tilde{\theta}_{y_{i}}}{\left\langle x_{i}-\overline{y}_{i}\right\rangle }\int\frac{du_{3}du_{4}d\theta_{3}^{u}d\theta_{4}^{u}}{\left\langle 34\right\rangle _{u}^{1+\frac{\lambda}{2}}}\frac{dv_{3}dv_{3}d\theta_{3}^{v}d\theta_{4}^{v}}{\left\langle 34\right\rangle _{v}^{1+\frac{\lambda}{2}}}\frac{\left\langle \Delta\overline{\phi}_{1}\,\Delta\phi_{2}\Phi_{3}\overline{\Phi}_{4}\prod_{i=1}^{n}\Phi\left(x_{i}\right)\Phi\left(y_{i}\right)\right\rangle }{\left\langle \Phi_{3}\overline{\Phi}_{4}\right\rangle \cdot\prod_{i=1}^{n}\left\langle \Phi\left(x_{i}\right)\Phi\left(y_{i}\right)\right\rangle }\;,
\end{split}
\end{equation}
$\phi$ being the bottom component of the superfield $\Phi$. In the integral we changed variables to the light-cone variables for the points $3,4$, and transformed the $2n$ integrals over $x_i,y_i$ to flat space. As a result, both sides of the ratio on the RHS can be calculated in flat space. We are interested in the behavior of the integrand at large $|u_i|,|v_i|$. Inside the Euclidean integrals, the integrand factorizes between the $u$'s and the $v$'s, and so we deal with each separately. Explicitly, the relevant terms in the ratio for the $u$-integral are:
\begin{equation}\label{eq:u_integrand}
\begin{split}
    &\left(\frac{\left\langle \Delta\overline{\phi}_{1}\,\Delta\phi_{2}\Phi_{3}\overline{\Phi}_{4}\prod_{i=1}^{n}\Phi\left(z_{i}\right)\Phi\left(w_{i}\right)\right\rangle }{\left\langle \Phi_{3}\overline{\Phi}_{4}\right\rangle }\right)_{u}
    =
    \left\langle 34\right\rangle _{u}^{\Delta}\left(\frac{\left(\text{...}\right)}{\left\langle 34\right\rangle _{u}}+\sum_{i,j}\frac{\left(\text{...}\right)}{\left\langle 3\,w_{i}\right\rangle \left\langle z_{j}\,4\right\rangle }\right)^{\Delta}\\
    &\qquad =
    \left\langle 34\right\rangle _{u}^{\Delta}\left(\frac{\left(\text{...}\right)}{u_{34}-2\theta_{3}^{u}\theta_{4}^{u}}
    +\sum_{i,j=1}^n\frac{\left(\text{...}\right)}{\left(u_{3}-w_{i}-2\theta_{3}^{u}\tilde{\theta}_{i}\right)\left(z_{j}-u_{4}-2\theta_{j}\theta_{4}^{u}\right)}\right)^{\Delta}\;.
\end{split}
\end{equation}
In the first equality we have plugged in the $n$-point function \eqref{eq:n_point_conjecture_main_text}, and separated the contraction of $3$ to $4$ from the rest of the contractions. The expressions ``$(...)$'' denote terms that depend only on the Euclidean (super)-coordinates.

We are interested in the large $u_3,u_4$ limit of this integral after performing the $\theta^u_3,\theta^u_4$ integrals. As the integrals acts as derivatives, we can consider each term separately. The Grassman integrals of either the $\langle 34\rangle_u^{\Delta-1-\frac{\lambda}{2}}$ term or the first term in the bracket multiply the bottom component of \eqref{eq:u_integrand} by $1/u_{34}$ (up to multiplicative constant). On the other hand, integrating one of the terms in the sum over $i,j$ multiplies the bottom component by a factor of $\tilde\theta_i/(u_3-w_i)$ from the $\theta_3^u$ integral, and $\theta_j/(z_j-u_4)$ from the $\theta_4^u$ integral. Note that the Euclidean Grassmann integrals won't change the over power in $u_3,u_4$. Together we have a factor of $1/(u_3 u_4)$, which is subleading at large $u_3,u_4$ compare to $1/u_{34}$. The $v$-integral is very similar:
\begin{equation}
\begin{split}
    &\left(\frac{\left\langle \Delta\overline{\phi}_{1}\,\Delta\phi_{2}\Phi_{3}\overline{\Phi}_{4}\prod_{i=1}^{n}\Phi\left(z_{i}\right)\Phi\left(w_{i}\right)\right\rangle }{\left\langle \Phi_{3}\overline{\Phi}_{4}\right\rangle }\right)_{v}
    =\left(\frac{\left\langle 34\right\rangle _{v}}{v_{3}v_{4}}\right)^{\Delta}\left(v_{3}v_{4}\frac{\left(...\right)}{\left\langle 34\right\rangle _{u}}+\sum_{i,j=1}^{n}\frac{\left(...\right)}{\overline{\left\langle 3w_{i}\right\rangle }\overline{\left\langle z_{j}4\right\rangle }}\right)^{\Delta}\\
	&\qquad=\left\langle 34\right\rangle _{u}^{\Delta}\left(\frac{\left(...\right)}{\left\langle 34\right\rangle _{u}}+\sum_{i,j=1}^{n}\frac{\left(...\right)}{\left(1-\bar{z}_{m}v_{3}-2\theta_{3}^{v}\bar{\tilde{\theta}}_{m}\right)\left(1-\bar{z}_{j}v_{4}-2\bar{\theta}_{j}\theta_{4}^{v}\right)}\right)^{\Delta}
\end{split}
\end{equation}
In terms of the overall power of $v_3,v_4$ the argument from the $u$-integrals is carried in the same way: the leading divergence multiplies the bottom component by $1/v_{34}$. The overall result is that at large $|u_i|,|v_i|$ the integral has the form
\begin{equation}
    \sim \int\frac{du_{3}du_{4}}{u_{34}^{1+\frac{\lambda}{2}}}\frac{dv_{3}dv_{4}}{v_{34}^{1+\frac{\lambda}{2}}}\cdot\frac{1}{u_{34}v_{34}},
\end{equation}
which diverges for $\lambda \le 0$. We note that the argument is not complete as we did not perform the Euclidean integrals, but we expect it to work just like the leading four-point integration \eqref{eq:4point_integral}.

We thus find that all the higher orders in $k_R$ also diverge only for $\lambda_L\leq 0$, and so our perturbative expansion is justified  and indeed the chaos exponent at $J=0$ is $\lambda_L(J\to 0^+)=0$. The transition into chaos should thus be continuous, as in figure \ref{fig:continuous_chaos}.

\subsection{Checking the validity of the approximation}\label{sec:validity_2}

Once again, we must now make sure that our approximation of neglecting $F_0$ in computing the chaos exponent is consistent. The analysis is similar to the one done for the disordered GFFs in section \ref{sec:validity_1}. In particular, we must read off the chaos exponent coming only from $F_0$ at $J=0$, denoted by $\lambda_L^{0}$. 

There is a simple trick for computing the analytically-continued four-point function required for $F_0$. Since the ``usual'' four-point function is a product of propagators (for the bottom component, see \eqref{eq:4_varphi}), we just have to replace each propagator with the corresponding analytically-continued propagator. The result is that $F_0$ is given by
\begin{equation}
    F_0\propto \frac{G_{lr,\Delta}(12)G_{lr,\Delta}(34)G_{R,\Delta}(14)G_{R,\Delta}(23)}{G_{lr,\Delta}(13)G_{lr,\Delta}(24)}+(3\leftrightarrow 4)\;.
\end{equation}
Here $G_{lr,\Delta}$ is the propagator between the different rails defined in \eqref{eq:Glr}, and 
\begin{equation}\label{eq:G_R}
    G_{R,\Delta}(1,2) = \frac{1}{\left( 4\sinh(\frac{t_{12}-x_{12}}{2})\sinh(\frac{t_{12}+x_{12}}{2}) \right)^\Delta}\;.
\end{equation}
At large $t_3\approx t_4=t$, $F_0$ behaves like $e^{0\cdot t}$, and so the chaos exponent predicted by $F_0$ is $\lambda_L^{0}=0$. Thus once again, the chaos exponent predicted by $F_0$ at $J=0$ is identical to the chaos exponent predicted by the kernel as we approach $J=0$ from above, $\lambda_L^{0}=\lambda_L(J\to 0)$, and so once again the chaos exponent is continuous at $J=0$ as discussed around equation \eqref{eq:continuity}.

\section{Conclusions}\label{sec:conclusions}

In this paper we discussed disorder around general CFTs, which allowed us to compute the chaos exponent $\lambda_L$ as a function of a continuous parameter $J$ in some specific models. We started by writing down a set of self-consistency equations for the two- and four-point functions (and also the OTOC) for a general disordered CFT. We then discussed models in which the disorder parameter $J$ is exactly marginal (at least at leading order in $1/N$). In principle this allowed us to compute the chaos exponent $\lambda_L(J)$ for any value of the disorder parameter $J$, and to follow the theory from weak to strong chaos. We managed to perform this analysis explicitly for disordered generalized free fields in $0+1$d and $1+1$d. In addition, we performed this analysis to leading order in $J$ in the disordered $\mathcal{N}=(2,2)$ $A_2$ minimal model. For the disordered generalized free fields we found a discontinuous transition into chaos, while for the disordered $A_2$ minimal model we found a continuous transition.

As discussed above, in principle the computation (at least at leading order in $J$) should be possible for all of the $A_{q-1}$ minimal models, since their four-point functions are known \cite{Mussardo:1988av}. It would be interesting to see whether the transitions to chaos in the general case would be continuous or discontinuous. In particular, the case $q=4$ should be the next simplest case after $q=3$, since it has central charge $c=3/2$ and is the $\mathcal{N}=(1,1)$ free chiral superfield \cite{Mussardo:1988av}. As a result, the computation should parallel the one done above for $q=3$, since we can map components of chiral superfields to vertex operators or fermionic operators, whose correlators are known. Naive dimensional analysis seems to indicate that the chaos exponent is continuous in this case, but the analysis should be done more carefully.

One can also extend this analysis to 2+1d. The Wess-Zumino models defined in equation \eqref{eq:IR_model} have an immediate generalization to 2+1d theories with $\mathcal{N}=2$ SUSY. In this case, only the $q=3$ case is a relevant deformation from the UV free field theory. This model was studied in \cite{Chang:2021fmd}, but it would be interesting to study the properties of the conformal manifold as well. This would be a much more daunting task than the $1+1$d theories discussed here, since the $1+1$d versions correspond to a disorder deformation of very simple CFTs (the minimal models), while the $2+1$d versions correspond to complicated CFTs.

An interesting result that was argued for above is the continuity relation \eqref{eq:continuity}. As a reminder, the chaos exponent near $J=0$ can either be computed explicitly by computing the OTOC at $J=0$ (and the result is denoted $\lambda_L^0$), or computing $\lambda_L(J)$ from the eigenvalue equation for the kernel $K_R$ and taking the limit $J\to 0$. Then we argued that $\lambda_L^{0}=\lambda_L(J\to 0)$. In addition, we showed that this relation is true in the explicit examples studied in this paper in sections \ref{sec:validity_1} and \ref{sec:validity_2}. It would be very interesting to understand whether this is a general result for disordered CFTs. If it is a general result, then in principle one could find whether the transition into chaos is continuous or discontinuous by performing a computation in a single copy of one core CFT (since computing $\lambda_L^0$ requires knowing only one copy of the core CFT). Then it would be interesting to understand precisely what set of conditions a core CFT is required to obey in order for the transition into chaos to be continuous or discontinuous.

Assuming this continuity condition on the chaos exponent, there are now many additional examples where the transition into chaos would be discontinuous - the idea would be to consider a single core CFT, and find $\lambda_L^{0}$ (see \eqref{eq:lambdaL0_def}). If the result is negative, then we expect a discontinuous transition into chaos, assuming that a conformal manifold can be constructed at leading order in $1/N$ as above. In particular, we expect that the non-SUSY minimal models will have a discontinuous transition using the results of \cite{Caputa:2016tgt}, and that the $2+1$d Ising model will have a discontinuous transition using the results of \cite{Caron-Huot:2020ouj}, if similar conformal manifolds exist.

An additional result is that the perturbative expansion for the chaos exponent was indeed consistent, which was nontrivial as explained in section \ref{sec:chaos_J0}. In particular, it is crucial that the integrals coming from higher-order corrections diverge at $\lambda$ which obeys $\lambda\leq \lambda_L^0$. Relatedly, one must show that equation \eqref{eq:ansatz} is obeyed. Indeed, once again we saw explicitly that this is obeyed in the specific examples discussed in this paper. It would be nice to prove this behavior for a general CFT.

Our self-consistency equations for the two- and four-point functions around a general disordered CFT can be useful outside of the scope of this paper. It would be interesting to see if there are additional core CFTs for which these equations can be solved exactly apart from free fields. This would be especially useful in cases where the disorder deformation is not exactly marginal, since then the equations would probably be solvable only in the limit $J\to\infty$ where conformal invariance may be restored, and so perturbation theory in $J$ will not be useful.
It would also be interesting to try to solve these equations in perturbation theory to high orders in $J$ in the case where the disorder is exactly marginal. Finally, more general correlators can also be computed for disordered free fields \cite{Gross:2017aos}, and it would be interesting to see if this is the case also for general disordered CFTs.

It would be interesting to understand the dependence of the chaos exponent on exactly marginal deformations also from a holographic perspective. In particular, \cite{Belin:2020nmp} studied a symmetric orbifold of the $\mathcal{N}=2$ SUSY minimal models of the type reviewed in section \ref{sec:disorder_and_conformal_manifolds}. These theories have a much smaller number of exactly marginal deformations, but far along the conformal manifold the authors found evidence for a (weakly curved) holographic dual. It would be interesting to try to apply the methods discussed above to these examples and study $\lambda_L$ as function of the exactly marginal deformation. If the theory is indeed holographic far away on the conformal manifold, $\lambda_L$ should reach its maximal value $\lambda_L = 1$. It would be interesting to show this and to study small deviations away from maximal chaos. 

Finally, it would be very interesting to compare our results with classical expectations for the onset of chaos. As discussed above, there are exact theorems which described the chaotic behavior of some models as they are deformed away from weak coupling, like the KAM theorem. There are incredibly useful tools in studying classical chaos, and hopefully a better understanding of the onset of quantum chaos will lead to similar tools.

\section*{Acknowledgements}

The authors would like to thank N. Brukner, C. Choi, R.R. Kalloor, B. Lian, J. Maldacena, O. Mamroud, M. Mezei, V. Rosenhaus, D. Tong and M. Watanabe for useful conversations.
We especially thank M. Rangamani and O. Aharony for interesting discussions and for comments on a draft of this paper.
The work was supported in part by an Israel Science Foundation center for excellence grant (grant number 2289/18), by grant no.2018068 from the United States-Israel Binational Science Foundation (BSF), and by the Minerva foundation with funding from the Federal German Ministry for Education and Research. MB is the incumbent of the Charles and David Wolfson Professorial Chair of Theoretical Physics. 

\newpage

\begin{appendix}

\addtocontents{toc}{\protect\setcounter{tocdepth}{1}}

\section{$\mathcal{N}=2$ SUSY conventions}\label{app:conventions}

Our conventions follow \cite{Bulycheva:2018qcp}. $\mathcal{N}=2$ superspace consists of a set of holomorphic and anti-holomorphic coordinates, which we call
\begin{equation}
Z=(z,\theta,\tilde{\theta})\;,\quad 
\bar Z=(\bar z, \bar\theta,\bar{\tilde{\theta}})\;.
\end{equation}
We also define superspace derivatives as
\begin{equation}
D=\frac{\partial}{\partial\theta}+\tilde{\theta}\frac{\partial}{\partial z}\;,\quad \bar D=\frac{\partial}{\partial\bar\theta}+\bar{\tilde{\theta}}\frac{\partial}{\partial \bar z}\;.
\end{equation}
Chiral superfields $\Phi$ then obey $D \Phi=\bar D \Phi=0$. We will denote anti-chiral superfields by $\bar \Phi$. A SUSY Lagrangian includes a Kahler potential and a superpotential and takes the form
\begin{equation}
\mathcal{L}=\int d^2\theta d^2\tilde{\theta} \Phi\bar{\Phi}+i\int d^2\theta W(\Phi)+i\int d^2\tilde{\theta}\bar W(\bar \Phi)\;.
\end{equation} 
Here we have defined $d^2\theta=d\theta d\bar{\theta}$ and $d^2\tilde{\theta}=d\tilde{\theta}d\bar{\tilde{\theta}}$. 

In $\mathcal{N}=2$ SCFTs, two-point functions of chiral operators are fixed:
\begin{equation}
\langle \bar{\Phi}(Z_1)\Phi(Z_2) \rangle=\frac{b}{\langle 12\rangle^\Delta\langle \bar 1\bar 2\rangle^\Delta}\;,
\end{equation}
where $b$ is a constant, $\Delta$ is the dimension of $\Phi$ and 
\begin{equation}
\langle 12\rangle=z_{12}-2\tilde{\theta}_1\theta_2-\theta_1\tilde{\theta}_1-\theta_2\tilde{\theta_2},\quad
\langle \bar 1\bar 2\rangle=\bar{z}_{12}-2\bar{\tilde{\theta}}_1\bar{\theta}_2-\bar{\theta}_1\bar{\tilde{\theta}}_1-\bar{\theta}_2\bar{\tilde{\theta}}_2\;.
\end{equation}
There are two superconformal cross-ratios:
\begin{equation}\label{eq:susy_ratio}
\chi_S=\frac{\langle 12\rangle \langle 34\rangle }{ \langle 14\rangle \langle 32\rangle },\quad
\bar{\chi}_S=\frac{\langle \bar 1\bar 2\rangle \langle \bar 3\bar 4\rangle }{ \langle \bar 1\bar 4\rangle \langle \bar 3\bar 2\rangle }\;.
\end{equation}

\section{Finding the subtracted $n$-point functions $n_s$ and $n_s'$}\label{app:ns_and_nsp}

Above we described how to define the subtracted $n$-point functions $n_s$ and $n_s'$ which appear in the SD equations for the two-point function and in the kernel for the four-point function respectively. In this appendix we perform an explicit example of an accelerated algorithm for generating these correlation functions. The algorithm is applied as follows. At order $n$, consider the full CFT $n$-point function. This can be decomposed in terms of connected $n$-point functions of lower order. We plug this decomposition into the corresponding diagram in the SD equations (kernel), and remove contributions which lead to disconnected parts. The remaining terms define $n_s$ ($n_s'$). We will do specific examples in this appendix and show that they match the previous definition.

\subsection{$n_s$ and the SD equations}\label{app:ns}

We start by doing specific examples by finding $4_s,6_s$ using the accelerated algorithm.

\begin{figure}
	\centering
	\begin{subfigure}[t]{1\textwidth}
		\centering
		\includegraphics[width=0.5\linewidth]{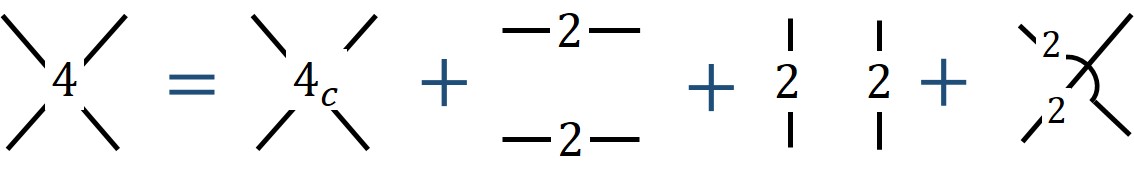}
		\caption{}
		\label{fig:algo4pta}
	\end{subfigure}%
	~ \\
	\begin{subfigure}[t]{1\textwidth}
		\centering
		\includegraphics[width=0.5\linewidth]{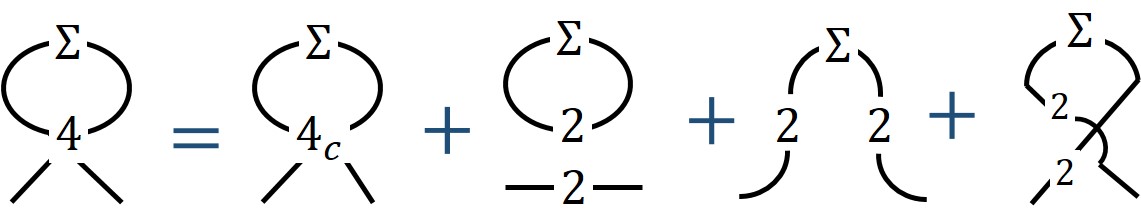}
		\caption{}
		\label{fig:algo4ptb}
	\end{subfigure}
	\caption{The algorithm for finding $4_s$. an $n$ corresponds to a full $n$-point function while an $n_c$ corresponds to a connected $n$-point function.}
	\label{fig:algo4pt}
\end{figure}

Let us start by performing this analysis explicitly for the four-point function $4_s$, whose contribution to the SD equations corresponds to contracting two of the legs with a $\Sigma$, see figure \ref{fig:sdeqs}. We start by decomposing the full CFT four-point function $4$ into fully connected $n$-point functions, see figure \ref{fig:algo4pta}. Next, we plug this decomposition into the contribution for the SD equations by contracting two of the legs with a $\Sigma$, see figure \ref{fig:algo4ptb}. It is clear that the second term on the RHS in figure \ref{fig:algo4ptb} has a disconnected component, and so we must remove it. Then $4_s$ is defined as taking the full four-point function \ref{fig:algo4pta} and subtracting the term in the decomposition which leads to a disconnected diagram, which is indeed the result of $4_s$ in figure \ref{fig:subtracted_n_points}.

\begin{figure}[]
	\centering
	\includegraphics[width=0.75\linewidth]{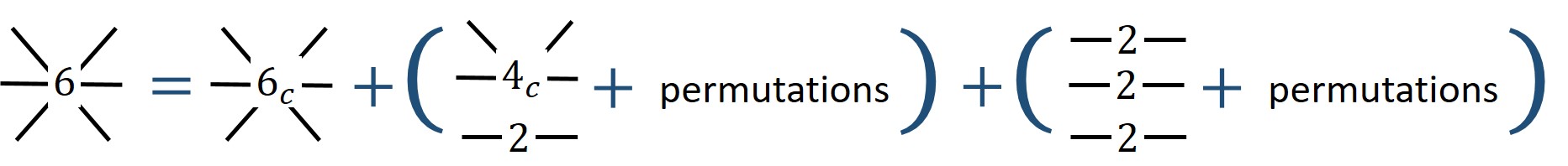}
	\caption{Decomposing the six-point function.}
	\label{fig:algorithm6pt}
\end{figure}

Next we do the analysis for the six-point function. The decomposition of the full six-point function into fully connected $n$-point functions is given in figure \ref{fig:algorithm6pt}. The corresponding contribution to the SD equations is obtained by contracting two pairs of external legs via a $\Sigma$, see figure \ref{fig:sdeqs}. Performing this contraction on each term in figure \ref{fig:algorithm6pt}, we find that again some of the diagrams lead to disconnected contributions. Subtracting these, we find $6_s$ shown in figure \ref{fig:subtracted_n_points}.

\subsection{$n_s'$ and the four-point function}\label{app:nsprime}

\begin{figure}[]
	\centering
	\includegraphics[width=0.75\linewidth]{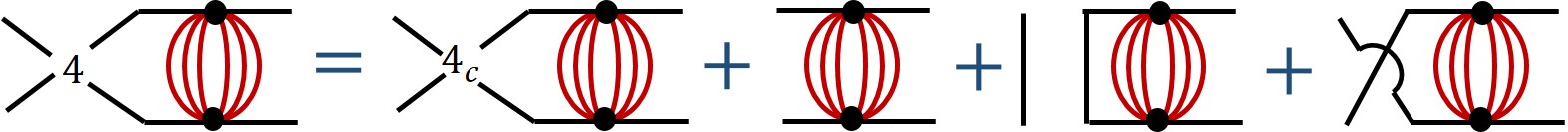}
	\caption{Decomposing the six-point function.}
	\label{fig:nsp}
\end{figure}

A similar method can be applied in order to find $n_s'$. We start by explicitly finding $4_s'$. The idea is the same as for the SD equation. We start with the full CFT four-point function $4$, and we decompose it into fully connected contributions as in figure \ref{fig:algo4pta}. We then plug this into the contribution of the $4_s'$ in the kernel in figure \ref{fig:kernel}, see figure \ref{fig:nsp}. We find again that there are disconnected contributions (specifically, the second diagram on the RHS), and they must be subtracted from the contribution of the four-point function. Then $4_s'$ is defined as the full four point function, with the diagram which leads to a disconnected contribution subtracted, as in figure \ref{fig:nsprime_examples}. A similar analysis for the 6-point function leads to the $6_s'$ defined in \ref{fig:nsprime_examples}.

\subsection{A consistency check: disordered free fields}\label{app:consistency}

As a consistency check, let us compare our SD equations in figure \ref{fig:sdeqs} to the standard result when expanding around a free field CFT, as in the SYK model in figure \ref{fig:freesdeqs}. We will do this up to order $J^6$. 

In the case where the core CFT is free, the $n$-point functions reduce to all of the possible ways of contracting the different legs using two-point functions, and the subtracted $n$-point function corresponds to removing contributions which lead to disconnected diagrams.
Let us see order-by-order that we recover the ``standard'' SD equation. At order $J^0$ this is obvious. At order $J^2$, the four-point function reduces to 3 possible contractions of the four operators, but one is removed due to the subtraction in $4_s$. As a result, we find two identical diagrams of the form
\begin{center}
	\includegraphics[width=0.13\linewidth]{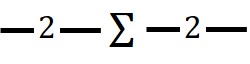}
\end{center}
The factor of two cancels with the factor of $1/2$ in \eqref{eq:G_order_J2}, and so we find the correct contribution.
Next consider order $J^6$. There are 15 ways of connecting the external legs of the 6-point function in pairs, but we subtract 7 of them in $6_s$. We are thus left with 8 diagrams of the form 
\begin{center}
	\includegraphics[width=0.2\linewidth]{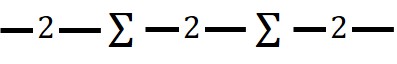}
\end{center}
Once again, the factor of $8$ cancels with the prefactor of $6_s$, and we are left with precisely the required contribution.

\section{The $A_2$ minimal model}\label{app:q3_min_model}

\subsection{Preliminaries}

In this section we study the $\mathcal{N}=2$ minimal model corresponding to the superpotential $W=X^3$, known as the $A_2$ minimal model. This minimal model has central charge $c=1$, and so it should correspond to the theory of a free compact boson $\phi\sim \phi+2\pi R$ at some value of the compactification radius $R$. 

It is slightly subtle to find the precise $c=1$ theory which the $X^3$ model maps to in the IR. The reason is that we are interested only in correlation functions in the CFT, and there are a handful of theories which differ only by gaugings of some discrete symmetries, so that correlators are invariant (assuming the operators are invariant under the symmetry we are gauging). In particular, the bosonic $c=1$ theory has four values of the radius $R$ where it has enhanced $\mathcal{N}=(2,2)$ SUSY, which are $R=\sqrt 3,\sqrt3/2$ and their T-dual values (see e.g.~\cite{Ginsparg:1988ui}). Indeed, the two theories at $R=\sqrt 3,\sqrt3/2$ are $\mathbb{Z}_2$ orbifolds of each other, and so correlation functions do not differ between them. It is thus not important which value of $R$ we choose for our purposes. 

However, this is not the end of the story. Our SUSY theory includes fermions, and so it requires a choice of spin structure, while the standard bosonic $c=1$ theory does not. Indeed, the theories at $R=\sqrt3,\sqrt3/2$ discussed above are theories with $\mathcal{N}=(2,2)$ SUSY, but where $(-1)^F$ has been gauged (as we will see later on). Instead, our SUSY theory should correspond to a ``fermionic''  $c=1$ theory, i.e.~a $c=1$ theory with a choice of spin structure.\footnote{We thank D. Tong for discussions on this issue.}
We will assume that the fermions can be introduced by an ``ungauging'' procedure of $(-1)^F$, i.e.~we will assume that there is some $\mathbb{Z}_2$ symmetry which upon gauging reintroduces the fermions. 

The bottom line is then that the theory with superpotential $W=X^3$ flows to an orbifold of the standard bosonic theory with $R=\sqrt 3/2$. Since orbifolds do not change correlators (as long as the operators are not charged under the symmetry), we may proceed for now focusing on the $R=\sqrt 3/2$ bosonic theory, and we will discuss explicitly the operators for which the orbifold is relevant.

Vertex operators for the theory with $R=\sqrt 3/2$ take the form
\begin{equation}
V_{n,m}=\exp\left( i\left(\frac{m}{2R}+n R  \right)\phi+ i\left(\frac{m}{2R}-n R  \right)\bar\phi\right)\;,
\end{equation}
with dimensions
\begin{equation}
    \begin{split}
        h&=\frac12\left(\frac{m}{2R}+n R \right)^2\;,\\
        \bar h&= \frac12\left(\frac{m}{2R}-n R \right)^2\;.
    \end{split}
\end{equation}
In these conventions, the dimension is $\Delta=h+\bar h$ and the spin is $\ell=|h-\bar h|$. 
In particular, there exist four vertex operators $V_{\pm2,0},V_{0,\pm 3}$ with dimensions $(h,\bar h)=(3/2,3/2)$, which is appropriate for an $\mathcal{N}=
(2,2)$ supersymmetric theory where $(-1)^F$ has been gauged.

We expect to find a chiral operator $X^{IR}$ of dimension $\Delta=1/3$ in this CFT. Expanding $X^{IR}$ in components as 
\begin{equation}
    X^{IR}=\varphi+\theta\psi+\bar{\theta}\bar{\psi}+\theta\bar{\theta}F \;,
\end{equation} we should be able to match each component with a vertex operator in the IR. Indeed, the vertex operator $V_{0,1}$ has dimensions $(h,\bar h)=(1/6,1/6)$, and so we match $\varphi\leftrightarrow V_{0,1}$ (and $\bar\varphi\leftrightarrow V_{0,-1}$). Next, we match $F\leftrightarrow V_{0,-2}$ and $\bar F\leftrightarrow V_{0,2}$, since they all have $(2/3,2/3)$.\footnote{To see that $F$ must have $m=-2$ and not $m=2$, we use the fact that $Q^2\varphi=F$, and the fact that we know the supersymmetry currents in terms of vertex operators (up to the subtelty of gauging $(-1)^F$ discussed below) and the OPE of two vertex operators. Specifically, computing the OPE of $J^2 \varphi$ and extracting the term which is proportional to $z^{-2}$ gives $F$.} 

Next we must find the fermion $\psi$ in terms of vertex operators. As discussed above, the bosonic $c=1$ CFT we are considering has $(-1)^F$ gauged, and so we should not be able to find $\psi$ in it. Instead, we are assuming that there is some ``ungauging'' procedure that allows us to reintroduce the fermions. In practice, this allows us to reintroduce the fermions and supercharges as vertex operators with non-integer values of $n,m$. We have checked that this reintroduction is consistent, in the sense that acting with the supercharges on the components of $X^{IR}$ give the expected results. These changes in the spectrum of the theory lead to a CFT which is not modular invariant, but this was expected due to the dependence on the spin structure (see e.g. \cite{Karch:2019lnn}).

Now we can compute $n$-point functions of $X^{IR}$ in the CFT. From now on we will ignore the IR superscript, and so $X$ is always understood to be the chiral operator of dimension $\Delta=1/3$ in the CFT. Since we know how to write the components of $X$ in terms of vertex operators, we now know how to compute and $n$-point function of them. 
To warm up, let us compute the two-point function of $X$. Superconformal invariance fixes \begin{equation}
    \langle \bar X X \rangle =\frac{1}{\left|\langle 12\rangle\right|^{2\Delta}}\;.
\end{equation} It is clear that the bottom component is precisely the $\varphi$ 2-point function $\frac{1}{|z|^{2\Delta}}$. Extracting the top component, we find that the two-point function of $F$ is $\frac{4\Delta^2}{\left|z_{12}\right|^{2\Delta+2}}$, which defines the normalization of $F$.

\subsection{4-point function}

Using the mapping of the components of $X$ to vertex operators in the $c=1$ free boson, we can immediately write down their 4-point functions:
\begin{align}
\langle \bar\varphi\varphi\bar\varphi\varphi \rangle &= \left|\frac{z_{13} z_{24}}{z_{12} z_{14}z_{23}z_{34}}\right|^{2\Delta}=\left|\frac{1}{z_{12} z_{34}}\right|^{2\Delta}|1-\chi|^{2\Delta}\label{eq:4_varphi}\\
\langle \bar FF\bar FF \rangle &=(4\Delta^2)^2 \left|\frac{z_{13} z_{24}}{z_{12} z_{14}z_{23}z_{34}}\right|^{8\Delta}=\left|\frac{1}{z_{12} z_{34}}\right|^{8\Delta}|1-\chi|^{8\Delta}\label{eq:4_F}\\
\langle \bar F F \bar \varphi\varphi \rangle & =4\Delta^2 \left|\frac{1}{z_{34}}\right|^{2\Delta}\left|\frac{1}{z_{12}}\right|^{8\Delta}\left|\frac{z_{13}z_{24}}{z_{14}z_{23}}\right|^{-4\Delta}=
\left|\frac{1}{z_{34}}\right|^{2\Delta}\left|\frac{1}{z_{12}}\right|^{8\Delta}\left|1-\chi\right|^{-4\Delta}\label{eq:2_varphi_2_F}
\end{align}
where $\Delta=1/3$. We have defined the conformal cross ratios
\begin{equation}
\chi = \frac{z_{12}z_{34}}{z_{14}z_{32}},\;\;\;\;\bar{\chi} = \frac{\bar{z}_{12}\bar{z}_{34}}{\bar{z}_{14}\bar{z}_{32}}\;. \end{equation}

We can now write down the full superspace 4-point function of the chiral operator $X$. 
In superspace there is a single superconformal ratio $\chi_S$, which is given by
\begin{equation}
\chi_S=\frac{\langle 12\rangle \langle 34\rangle }{ \langle 14\rangle \langle 32\rangle }
\end{equation}
where
\begin{equation}
\langle 12\rangle = z_{12}-2\tilde \theta_1\theta_2-\theta_1\tilde\theta_1 -\theta_2\tilde\theta_2
\end{equation}
and it is easy to find the 4-point function of $X$ in superspace using the results for its components:
\begin{equation}\label{eq:4_X_app}
 \langle \bar X X \bar X X \rangle =\left|\frac{1}{\langle 12\rangle\langle 34\rangle}\right|^{2\Delta}|1-\chi_S|^{2\Delta}\;.
\end{equation}
We have checked that the components of this 4-point function match our expectations (including the fermionic components).

\subsection{Higher $n$-point functions}

We now conjecture the general superspace form for the correlation function of $2n$ $X$'s of the form $\langle X(x_1)...X(x_n)\bar X(y_1)...\bar X(y_n)\rangle$, and provide some nontrivial consistency checks for it. 

Our conjecture for the $2n$-point function is
\begin{equation}\label{eq:n_point_conjecture}
\langle X(x_1)...X(x_n)\bar X(y_1)...\bar X(y_n)\rangle=\left|\sum_{\sigma\in S_n} \text{sign}(\sigma)\prod_{i=1}^n 
\frac{1}{\langle x_i, y_{\sigma(i)}\rangle}\right|^{2\Delta}\;.
\end{equation}
This can be written in a more concise form as
\begin{equation}
\left| \det C(x_i,y_j) \right|^{2\Delta}
\end{equation}
where $C$ is a variant of the Cauchy matrix:
\begin{equation}
C=\left( \begin{array}{cccc}
\frac{1}{\langle x_1 ,y_1 \rangle} & \frac{1}{\langle x_1 ,y_2  \rangle} & \cdots & \frac{1}{\langle x_1 ,y_n  \rangle} \\
\frac{1}{\langle x_2 ,y_1  \rangle} & \frac{1}{\langle x_2 ,y_2 \rangle } & \cdots & \frac{1}{\langle x_2 ,y_n  \rangle} \\
\vdots & \vdots & \ddots & \vdots \\
\frac{1}{\langle x_n ,y_1  \rangle} & \frac{1}{\langle x_n ,y_2 \rangle } & \cdots & \frac{1}{\langle x_n ,y_n \rangle}
\end{array}\right)\;.
\end{equation}

We now describe some consistency checks on this result. First, it is symmetric under a permutation of the $x$'s and of the $y$'s. Next, we have checked explicitly for $n=1,2,3$ (i.e.~for the 2,4 and 6-point functions) that this result is correct by explicitly comparing to the expected result using the different components of $X$. Next, it is easy to check that at least the bottom component (with all Grassmanian coordinates set to zero) takes the correct value for any $n$. To see this, note that we expect the result to be
\begin{equation}
	\left| \frac{\prod_{i<j}^n x_{ij}y_{ij}}{\prod_{i,j}^n (x_i-y_j)}\right|^{2\Delta}\;.
\end{equation}
This matches the bottom component of \eqref{eq:n_point_conjecture} once we use the Cauchy determinant formula:
\begin{equation}
\langle X(x_1)...X(x_n)\bar X(y_1)...\bar X(y_n)\rangle|_{\text{b}}=\sum_{\sigma\in S_n} \text{sign}(\sigma)\prod_{i=1}^n \frac{1}{ x_i-y_{\sigma(i)}}= \frac{ \prod_{i<j }^n x_{ji}y_{ij}}{\prod_{i,j }^n(x_i-y_j)}\;.
\end{equation}
Where $|_{\text{b}}$ denotes taking the bottom component of the expression. Finally, we can also check that the components proportional to $\tilde\theta\theta$ match. To see this, we focus on the holomorphic part of the $2n$-point function, and calculate $\langle F\varphi^{n-1}\bar F\bar\varphi^{n-1}\rangle$. This corrsponds to computing
\begin{equation}
	\frac{d}{d\tilde\theta_{x_1}\theta_{y_1}}\langle (X\bar X)^n\rangle|_{\text{b}}\;.
\end{equation}
We are using the fact that at this order, we only need to take into account cases where $\tilde\theta_{x_1},\theta_{y_1}$ appear in the combination $\tilde\theta_{x_1}\theta_{y_1}$, which will not be true with higher derivatives. In this case we can use the general formula for the derivative of a determinant:
\begin{equation}
\frac{d}{dt}\det A = \det A \Tr (A^{-1}\frac{d}{dt}A)\;.
\end{equation}
We will not be able to use this formula to prove that this is the correct form for any component, since there will be various minus signs from the ordering of the $\theta$'s. But for this component there will be no sign problems, since we are taking the derivative $\frac{d}{d\tilde\theta_{x_1}\theta_{y_1}}$ and $\tilde\theta_{x_1},\theta_{y_1}$ always appear in the same order. Using this formula, if we take $\frac{d}{d\tilde\theta_{x_1}\theta_{y_1}}$ of the holomorphic part of our $2n$-point function and take the bottom component, this is the same as computing
\begin{equation}
\Delta \det(C)^{\Delta}\Tr (C^{-1}\frac{d}{dt}C)\;.
\end{equation}
Let us compute this. The elements of our matrix are of the form
\begin{equation}
C_{ij}=\frac{1}{\left\langle x_i ,y_{\sigma(i)}\right\rangle }= \frac{1+\frac{2\tilde\theta_{x_i }\theta_{y_{\sigma(i)}}}{x_i -y_{\sigma(i)}}}{x_i -y_{\sigma(i)}}
\end{equation}
and so $\frac{d}{d\tilde\theta_{x_1}\theta_{y_1}}C_{ij}|_{\theta=0}=2\frac{1}{(x_1 -y_1)^2}\delta_{i1}\delta_{j1}$. The inverse of the Cauchy matrix is also known, it is
$$
C^{-1}_{i j}=\frac{
	\prod_{k=1}^{n}\left(x_{j}-y_{k}\right)\left(x_{k}-y_i\right)}
{\left(x_{j}-y_i\right)\left(\prod_{1 \leq k \leq n \atop k \neq j}\left(x_{j}-x_{k}\right)\right)\left(\prod_{1 \leq k \leq n \atop k \neq i}\left(-y_i+y_{k}\right)\right)}\;.
$$
Putting these together we find that $\frac{d}{d\tilde\theta_{x_1}}\frac{d}{d\theta_{y_1}}|_{\theta=0}$ of the holomorphic part of our $2n$-point function is 
\begin{equation}
2\Delta \det(C)^{\Delta} C^{-1}_{11}\frac{1}{(x_i -y_{\sigma(i)})^2}\;.
\end{equation}
Explicitly, this is equal to
\begin{equation}
2\Delta \left( \frac{ \prod_{i<j}^nx_{ji}y_{ij}}{\prod_{i,j}(x_i-y_j)}\right)^{\Delta} \frac{
	\prod_{k=1}^{n}(x_1-y_{k})(x_{k}-y_1)}
{\left(x_1-y_1\right)\left(\prod_{2 \leq k \leq n}(x_{1k})(-y_{1k})\right)}\frac{1}{(x_1 -y_1)^2}\;.
\end{equation}
Separating the terms which include $i,j=1$ and the rest of the terms and setting $\Delta=1/3$ we find
\begin{equation}
\frac{d}{d\tilde\theta_{x_1}\theta_{y_1}}\langle (X\bar X)^n\rangle|_{hol,\theta=0}=2\Delta \left( \frac{ \prod_{1<i<j}^n x_{ji}y_{ij}}{\prod_{i,j\neq 1}(x_i-y_j)}\right)^{\Delta} \left(\frac{
	\prod_{k=2}^{n}(x_1-y_{k})(x_{k}-y_1)}
{\prod_{k=2}^n x_{k1}y_{1k}}\right)^{2\Delta}\frac{1}{(x_1 -y_1)^{4\Delta}}\;.
\end{equation}
Adding the anti-holomorphic part of this correlator in the same way, we find exactly the expected result for $\langle|F|^2|\varphi^{n-1}|^2\rangle$. Since we only considered the holomorphic part, the same calculation also shows that we get the correct result for $\langle\psi^2|\varphi^{n-1}|^2\rangle$.

\end{appendix}


\bibliographystyle{JHEP}
\bibliography{refs}

\end{document}